\documentclass[fleqn,usenatbib]{mnras}

\usepackage{newtxtext,newtxmath}

\usepackage[T1]{fontenc}
\usepackage{ae,aecompl}

\usepackage{graphicx}	
\usepackage{amsmath}	
\usepackage{enumerate}
\usepackage{enumitem}
\usepackage{siunitx}
\usepackage{CJK}

\newcommand{\Sref}[1]{Section \ref{#1}}
\newcommand{\Tref}[1]{Table \ref{#1}}

\sfcode`\.=1001\sfcode`\?=1001\sfcode`\!=1001
\newcommand{\Fref}[1]{\ifhmode \ifnum\spacefactor=1001 Figure \ref{#1}\else Fig.\ \ref{#1}\fi \else Figure \ref{#1}\fi}
\newcommand{\Eref}[1]{\ifhmode \ifnum\spacefactor=1001 Equation (\ref{#1})\else equation (\ref{#1})\fi \else Equation (\ref{#1})\fi}
\newcommand{\Teffnom}{\TextOrMath{$T^_{\mathrm{eff}\odot}$}{T_{\mathrm{eff}\odot}}}
\newcommand{\Teff}{\ensuremath{T_{\mathrm{eff}}}}
\DeclareSIUnit\angstrom{\mbox{\normalfont\AA}}
 
\title[SDST I: EPIC stellar parameters]{Survey for Distant Solar Twins (SDST) -- I. EPIC method for stellar parameter measurement}

\author[C. Lehmann et al.]{
Christian Lehmann,$^{1}$\thanks{E-mail: clehmann@swin.edu.au}
Michael T. Murphy,$^{1}$
Fan Liu (刘凡),$^{1}$
Chris Flynn,$^{1,2}$
and Daniel A. Berke$^{1}$
\\
$^{1}$Centre for Astrophysics and Supercomputing, Swinburne University of Technology, Hawthorn, Victoria 3122, Australia 
\\
$^{2}$ARC Centre of Excellence for Gravitational Wave Discovery (OzGrav), Australia
\\
}

\date{Accepted XXX. Received YYY; in original form ZZZ}

\pubyear{2022}

\begin{document}
\begin{CJK*}{UTF8}{gbsn}
\label{firstpage}
\pagerange{\pageref{firstpage}--\pageref{lastpage}}
\maketitle

\begin{abstract}
Solar twins are stars of key importance to the field of astronomy and offer a multitude of science applications. Only a small number ($\lesssim200$) of solar twins are known today, all of which are relatively close to our Sun ($\lesssim\SI{800}{pc}$). 
The goal of our Survey for Distant Solar Twins (SDST) is to identify many more solar twin and solar analogue stars out to much larger distances ($\sim\SI{4}{kpc}$). In this paper, we present a new method to identify solar twins using relatively low $S/N$, medium resolving power ($R\sim 28{,}000$) spectra that will be typical of such distant targets observed with HERMES on the $\SI{3.9}{\metre}$ Anglo-Australian Telescope (AAT).
We developed a novel approach, namely EPIC, to measure stellar parameters which we use to identify stars similar to our Sun.
EPIC determines the stellar atmospheric parameters (effective temperature $\Teff$, surface gravity $\log g$ and metallicity [Fe/H]) using differential equivalent width (EW) measurements of selected spectroscopic absorption features and a simple model, trained on previously analysed spectra, that connects these EWs to the stellar parameters. The reference for the EW measurements is a high $S/N$ solar spectrum which is used to minimise several systematic effects.
EPIC is fast, optimised for Sun-like stars and yields stellar parameter measurements with small enough uncertainties to enable spectroscopic identification of solar twin and analogue stars up to $\sim\SI{4}{kpc}$ away using AAT/HERMES, i.e.\ $\sigma\left(\Teff, \log g, \textrm{[Fe/H]}\right) = \left(\SI{50}{\kelvin, \SI{0.08}{dex}, \SI{0.03}{dex}}\right)$ on average at $S/N=25$.
\end{abstract}




\section{Introduction}\label{sec:Intro}
Solar twins are stars which are as intrinsically similar to the Sun as possible. 
Searches for solar twins have a history dating back to \citet[][]{Hardorp1978}, and have utilised a range of photometric and spectroscopic techniques.
Solar twins have proven to be excellent probes of astrophysics in our Galaxy because they allow us to reference other stars to the Sun, which is otherwise impossible because the Sun is too bright.
For example, solar twins have been used to model physical stellar characteristics like mass, age, chemical composition and the mixing-length parameter \citep{Bazot_2018}, track the chemical and kinematic evolution of the Galaxy \citep{Nissen2015, Nissen2016, Liu2016, Liu2019, Lorenzo-Oliveira2019, Carlos2020}, study the behaviour of optical lines with respect to stellar activity \citep{Flores2018}, explore the relation between age and magnetic activity \citep{LorenzoOliveira2018}, and the chemical implication of exoplanets on their host stars \citep{Melendez2009, Ramirez2009, Botelho2018, Liu2020}. They also provide an alternative means of wavelength calibrating high-resolution spectrographs \citep[e.g.][]{2015MNRAS.447..446W}.

These applications motivated many past searches for solar twins, which have been conducted with a range of methods.
\citet[][]{Hardorp1978} searched for Sun-like stars to study the metal content of the Sun compared to similar stars. They analysed the ultraviolet spectrum of solar type stars (G2V) in segments of $\SI{20}{\angstrom}$ which they normalised with solar spectra (e.g.\ reflected by asteroids, planetary satellites, etc.). The ultra-violet spectrum is sensitive to changes in temperature and element abundances which makes it a good spectral range to predict how Sun-like candidate stars are.
\citet{CayreldeStrobel1981}, \citet{CayreldeStrobel1989}, \citet{Friel1993} and \citet{CayreldeStrobelG.1996SrtS} continued the search for solar twins. They checked spectra of photometrically identified solar twins to examine the solar twin classification with a more directly measured temperature.
Other studies \citep[e.g.][]{King2005, Melendez2006, Melendez2012, Galarza2016, Bedell2017, Galarza2021} spectroscopically analysed and confirmed new solar twins ($\sim20$ in total) using the stellar atmospheric parameters ($\Teff, \log g$ and [Fe/H]) to determine how Sun-like their twin candidates were. They achieved uncertainties as low as $\sigma\left(\Teff, \log g, \textrm{[Fe/H]}\right) = \left(\SI{15}{\kelvin, \SI{0.03}{dex}, \SI{0.01}{dex}}\right)$ for these parameters. This precision is needed to achieve specific science goals with solar twins, e.g.\ to search for stars with the potential to host planetary systems similar to ours or to explore odd-even effects in abundance patterns with respect to solar abundances.
\citet[][]{Datson2012, Datson2014, Datson2015} identified solar twins using a differential method with a solar spectrum as the reference. They measured EW differences between spectra of candidate stars and the solar spectrum to identify solar twins. Their results inspired the differential approach that is used in this work.

The motivation for our Survey for Distant Solar Twins (SDST) is to discover solar twins as probes for possible variations in the strength of electromagnetism -- the fine-structure constant $\alpha$ -- across our Galaxy (Murphy et al. in prep; Berke et al. 2022a, b; in prep).
Stars closer to the Galactic centre sample regions of higher dark matter density, and so our project aims to test for any fundamental, beyond-Standard-Model connections between electromagnetism and dark matter.
Such connections are proposed within some theoretical models of fundamental physics \citep[e.g.][]{Olive2002,Eichhorn2018,Davoudiasl2019}.
However, all currently known solar twins are close to our solar system (within 800\,pc) because more distant stars would be too faint to have been useful for other astrophysical studies.
This raises a key challenge: the need to identify and spectroscopically confirm solar twins at large distances $\ga 800$\,pc, i.e.\ with magnitudes $V\ga15$\,mag.

Berke et al. (2022a; in prep) and Murphy et al. (in prep) recently demonstrated that solar twins can be used to constrain variation in $\alpha$, at a level of precision two orders of magnitude better than current astronomical tests. Their method is measuring absorption line separations with high precision to constrain the variation of $\alpha$ between target stars. 
Solar twins are ideal for this purpose as we can compare them both with each other, the Sun and local, close-by solar twins without introducing additional systematic effects. We use the stellar atmospheric parameters, $\Teff$, [Fe/H] and $\log g$, together as a proxy to determine how Sun-like a solar twin is.

We adopt the following definitions of solar twins and solar analogues from Berke et al. (2022b; in prep):
\begin{align}
    \text{Solar twin} &= \left\{
    \begin{array}{r@{\,}l}\label{eq:ST_def}
         \Teffnom&\pm\,\SI{100}{\kelvin}, \\
         \log{g_\odot}&\pm\,\SI{0.2}{dex},\\
         \mathrm{[Fe/H]_\odot}&\pm\,\SI{0.1}{dex},\\
    \end{array} \right. \\
    \text{Solar analogue} &= \left\{
    \begin{array}{r@{\,}l}\label{eq:SA_def}
        \Teffnom&\pm\,\SI{300}{\kelvin},\\
        \log{g_\odot}&\pm\,\SI{0.4}{dex},\\
        \mathrm{[Fe/H]_\odot}&\pm\,\SI{0.3}{dex}\\
    \end{array} \right.
\end{align}
where we use the following values for solar stellar parameters: $(\Teffnom, \log g_\odot$, [Fe/H]$_\odot) = (\SI{5772}{\kelvin}, \SI{4.44}{dex}, \SI{0.0}{dex})$ \citep[][]{Prsa2016}.
Note that these definitions do differ in the literature, e.g.\ \citet{CayreldeStrobel1981, CayreldeStrobel1989, Friel1993, CayreldeStrobelG.1996SrtS} defined solar twins using the same three stellar parameters above but with narrower ranges as well as restrictions on other physical parameters like age, microturbulence, chemical composition, etc. Such narrow definitions are not practical for our search because of how unlikely it is to find such a perfect match to the Sun.
For the same reason, other studies \citep[e.g.][]{Mello2014} defined the term `solar analogue' as stars that are less Sun-like than solar twins and therefore easier to identify as well as more practical.
Berke et al. (2022b; in prep.) demonstrated that solar analogues, as defined in \Eref{eq:SA_def}, are reliable probes of variations in the fine-structure constant between stars, so they are the prime targets of our survey.

The goal of the SDST is to identify a large number of solar analogues (between 200 and 300, after initial photometric pre-selection of $\sim$600), up to $\sim \SI{4}{kpc}$ away, using medium-resolution spectroscopy. The $\sim$40--50 most Sun-like of these, spread across several distance bins, can then be followed-up with high-precision spectrographs to provide accurate measurements of the fine-structure constant. The High Efficiency and Resolution Multi-Element Spectrograph (HERMES) at the $\SI{3.9}{\metre}$ Anglo-Australian Telescope (AAT) is ideal for the purpose of spectroscopically confirming solar analogue candidates as it can obtain 392 spectra simultaneously at a moderate resolving power ($R\approx28{,}000$). Nevertheless, Sun-like stars at $\SI{4}{kpc}$ will only result in $S/N\sim5$ per pixel spectra in $\SI{1}{\hour}$ of observation. We will combine HERMES spectra from consecutive nights to improve the $S/N$ of our distant targets to $\sim25$ per pixel.
Therefore, our aim was to develop a method to use these low $S/N$ combined HERMES spectra to confirm distant solar analogues with a precision of $\sigma\left(\Teff, \log g, \textrm{[Fe/H]}\right) \la \left(\SI{100}{\kelvin, \SI{0.2}{dex}, \SI{0.1}{dex}}\right)$.

We present a novel approach to identify solar twins and analogues optimised to work with HERMES spectra. We model the EWs of spectral absorption features as functions of their stellar parameters.
The method has two major advantages: 
\begin{enumerate}[labelwidth=*, leftmargin=*]
    \item It leverages existing spectral data from the Galactic Archaeology with HERMES (GALAH) survey \citep{Martell2017, Buder2018} to create models for absorption features. The library of stacked spectra from \citet{Zwitter2018} provides sample spectra with high $S/N$ (if not otherwise specified we use signal-to-noise per pixel within this work) and HERMES resolution on a grid of varying stellar parameters, which is ideal to create this model.
    \item The method operates fully differentially with respect to a high precision solar spectrum \citep[][]{Chance2010}. Specifically, the reference spectrum is used to homogenise the processes of continuum normalisation and radial velocity correction on a line-by-line basis. This ensures that a line's EW is measured in the same way for all target spectra and, therefore, differences between stars can be reliably traced.
\end{enumerate}
These advantages make it possible to measure the stellar parameters $\Teff$, [Fe/H] and $\log g$ with low uncertainties while also minimising computation time, i.e.\ $\sigma\left(\Teff, \log g, \textrm{[Fe/H]}\right) \approx \left(\SI{50}{\kelvin, \SI{0.08}{dex}, \SI{0.03}{dex}}\right)$ for $S/N=25$ per $\SI{0.063}{\angstrom}$ pixel in $\la \SI{10}{\second}$ per star.

This paper is structured as follows. In \Sref{sec:Method} we present our algorithm for solar analogue identification in detail. In \Sref{sec:Results} we describe the results of our method when applied to the solar reference spectrum \citep[][]{Chance2010} as well as spectral data from AAT/HERMES and HARPS at the ESO $\SI{3.6}{\metre}$ telescope in La Silla, Chile. In \Sref{sec:discussion} we discuss the impact of this method on the project to map out the fine-structure constant $\alpha$ throughout the Milky Way and compare our method to other stellar parameter measurement algorithms in the literature. \Sref{sec:conclusion} concludes and summarises the paper.

\section{EPIC stellar parameter measurement}\label{sec:Method}
\begin{figure}
	\centering
	\includegraphics[width=0.475\textwidth]{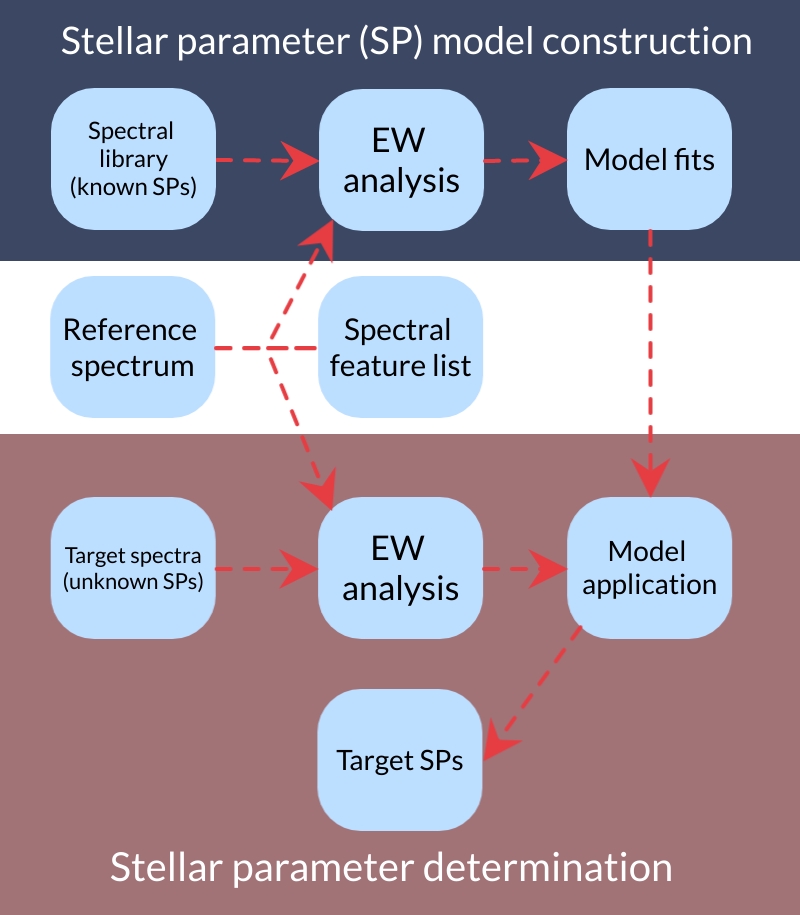}
	\caption{Schematic of overall approach used in EPIC to measure stellar parameters (SP) of target star spectra using differential equivalent width (EW) measurements. The white central area contains the essential reference inputs on which the method relies, while the blue area represents the SP model construction mode and the red area the SP determination mode. We explain this approach in detail in \Sref{sec:Method}.}
	\label{fig:Flowchart}
\end{figure}
We developed the `EPIC' (\textbf{E}W com\textbf{P}ar\textbf{I}son \textbf{C}ode) algorithm to determine stellar parameters in low $S/N\sim25$ spectra of solar analogue stars (henceforth `target spectra') with moderate resolving power ($20{,}000\lesssim R\lesssim28{,}000$). The EPIC code is publicly available online in \citet{CLehmann942022}.
EPIC analyses a predetermined set of absorption features (\Sref{sec:linelist}) to measure the stellar atmospheric parameters, i.e.\ effective temperature $\Teff$, surface gravity $\log g$ and metallicity [Fe/H], in target stars. 
We utilise two techniques to minimise uncertainties and improve efficiency in this method: (i) We use a set of high-quality library spectra to determine the sensitivity of absorption line EWs to the stellar parameters. (ii) We measure the EWs in a fully differential manner with respect to a high-quality reference spectrum with well-known stellar parameters.

\Fref{fig:Flowchart} visualises the overall approach which we split into two different operation modes:
\begin{enumerate}[labelwidth=*, leftmargin=*]
    \item \textbf{Stellar parameter model construction:}
    EPIC measures the EWs of a list of absorption features in a set of library spectra. We determine the difference in EW with respect to a high $S/N$ reference spectrum. The variation of EW differences in each feature is modelled as a simple polynomial function of the three stellar atmospheric parameters (\Sref{sec:stellar_par}).
    The library must comprise high $S/N$ spectra with similar resolving power as the target spectra and must have precisely determined stellar parameters.
    \item \textbf{Stellar parameter determination:} Using the model established in (i), EPIC determines the stellar parameters of target spectra. The EWs are measured with respect to the same reference spectrum used for model construction. EPIC then finds the best fit of the individual target EWs to the model to determine stellar parameters.
\end{enumerate}

The precision for measured stellar parameters depends on several factors: the $S/N$ of the target spectrum, systematic errors in the EWs, how well the model fits the library and target EW measurements, as well as the choice of spectral features in the line list. We focused considerable effort on addressing these possible origins of uncertainty and emphasise them within the detailed method explanation below.

While the outlined approach could, in principle, be applied to any spectrum from a wide range of different instruments, we focus on HERMES in this paper. HERMES \citep[][]{Sheinis2015} is a multi-object spectrograph mounted on the $\SI{3.9}{\metre}$ AAT, fed by 392 deployable fibres via the 2-degree-field (2dF) unit. It achieves moderate resolving power ($R\sim28{,}000$) in four relatively narrow wavelength bands, each recorded in a different arm of the instrument: 
\begin{itemize}[labelwidth=*, leftmargin=*]
    \item Blue (B): 4713--4901\,\AA
    \item Green (V): 5649--5872\,\AA
    \item Red (R): 6478--6736\,\AA
    \item Infra-Red (IR): 7585--7885\,\AA.
\end{itemize}
The combination of mirror size, multiplexing and resolving power allow the acquisition of $S/N \ga 25$ (in the red CCD) spectra of Sun-like stars at distances up to $\sim\SI{4}{kpc}$ which should enable us to measure spectral stellar parameters with adequate precision.
Furthermore, a crucial advantage of HERMES for identifying solar analogues at such large distances is the availability of the spectral library \citep{Zwitter2018} established by the GALAH survey \citep{Martell2017}. They stacked $336{,}215$ GALAH spectra into bins spanning a wide range of the three stellar atmospheric parameters.
These stacked spectra have high $S/N$ so that, for the purposes of this work, they are effectively noiseless and have a resolving power almost identical to (though slightly lower than) individual HERMES spectra. This spectral library makes it possible to establish a model connecting differences in EWs between absorption features in reference and target spectra to stellar parameters.
In this section, we describe our application of the EPIC approach to HERMES spectra.

\subsection{Preparing spectra for EPIC}\label{sec:preparing}
We prepare different spectral datasets so that the EPIC algorithm can be applied. Several different input spectra are required: a reference spectrum, target spectra, and stacked (i.e.\ library) spectra. Additionally, we created an algorithm which can prepare non-HERMES high resolution spectra so that they can be analysed by EPIC as if they were HERMES spectra. We describe below how each type of spectrum is prepared for use in EPIC.

\subsubsection{Reference spectrum}\label{sec:Ref_spec}
A high $S/N$, high-resolution spectrum is used as a reference. When searching for solar analogues, we utilise the solar atlas (`KPNO2010') from \citet[][]{Chance2010} as our default reference spectrum. It covers all four HERMES bands (wavelength range of 2990--10010\,\AA), is effectively noiseless for our use, has a high resolving power ($R = 200{,}000$--$300{,}000$) and is corrected for telluric absorption.

We next prepare the reference spectrum for comparison with the HERMES data.
The first step is to cut it into the four wavelength ranges (also called `bands') that are covered by HERMES. We include a $\SI{20}{\angstrom}$ wider wavelength range on both ends of each band as a buffer to avoid edge effects in subsequent steps, i.e.\ the convolution to HERMES resolution.
We then project the wavelength grid from the original spectrum onto a new version with constant wavelength spacing using the SpectRes \citep{Carnall2017} Python package. The wavelength grid is still very fine after this step compared to HERMES spectra (pixel-width $\sim \SI{0.01}{\angstrom}$).
\begin{figure}
	\centering
	\includegraphics[width=0.475\textwidth]{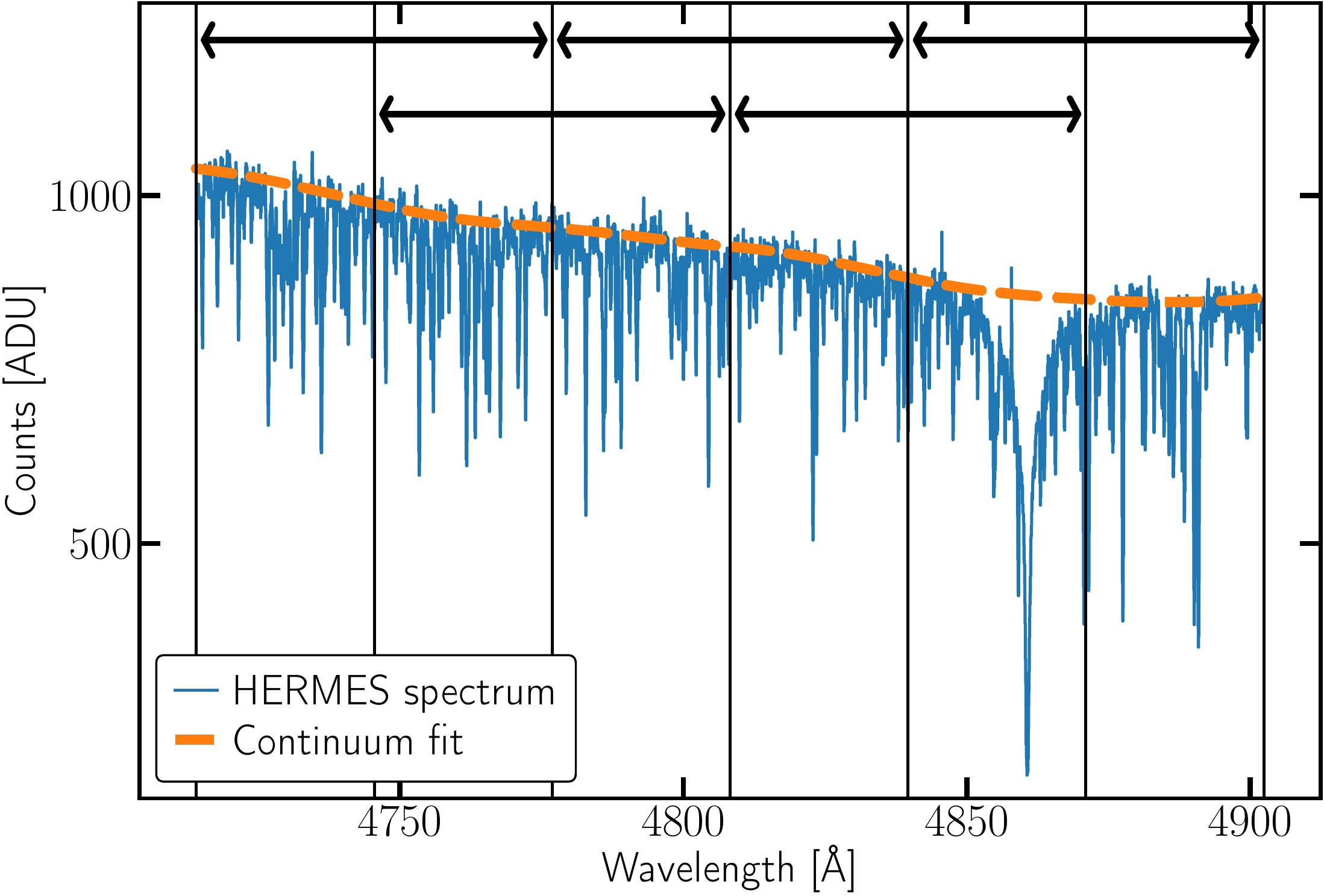}
	\caption{B-band in a representative high-$S/N$ HERMES spectrum from the GALAH survey. The black arrows mark the overlapping sub-spectra that we derive the normalisation from. The orange dashed line shows the resulting continuum after the local sub-continua are combined.}
	\label{fig:norm_split}
\end{figure}

Next, we normalise the spectrum by identifying continuum regions and fitting a polynomial of flux versus wavelength.
The first part of the normalisation process is to cut each band into overlapping sub-spectra (\Fref{fig:norm_split} illustrates this step). This is done to simplify the normalisation as we can now calculate the local continuum on each sub-spectrum independently.
We mask out the hydrogen absorption ($4855\leq \lambda \leq \SI{4870}{\angstrom}$ and $6555\leq \lambda \leq \SI{6575}{\angstrom}$) and strong telluric features ($7587\leq \lambda \leq \SI{7689}{\angstrom}$) because the algorithm would identify pixels within these wide emission/absorption areas as part of the continuum. We also mask any pixels that are more than 2.5 times the median flux in each selected sub-spectrum to initially select against strong cosmic rays (this is relevant when normalising non-reference spectra; see \Sref{sec:tar_prep}).
We remove pixels with the lowest 75\% of fluxes to select against absorption lines in each sub-spectrum and fit a third-order Legendre polynomial to those that remain. We calculate the standard deviation of these fluxes from the fit and select against those $>3$ standard deviations above and $>1.5$ standard deviations below the fit. The fit and deselection process is repeated until the number of pixels remains constant.
The final fit represents the calculated continuum of the sub-spectrum. 
We join the sub-continua together by weighting each sub-continuum linearly in the overlapping regions; the continuum closest to the centre of an individual sub-spectrum has close to 100\% weight while the edges have close to 0\% weight. 
This creates a continuum approximation for the full wavelength band. We normalise by dividing each pixel by its respective continuum value.

Note that, at the end of the reference spectrum preparation, the resolving power is not reduced as the desired resolving power is a function of both the target spectrum and the wavelength (\Sref{sec:measurement_EW}). Furthermore, if a different reference spectrum is used in EPIC, a different stellar parameter model (\Sref{sec:stellar_par}) must be created. This is because the EW analysis is differential, and hence dependent on the reference used.

\subsubsection{Target spectra}\label{sec:tar_prep}
The target stars observed with HERMES are the main focus of this work. To discover new, distant solar analogues, we need to use HERMES spectra with $S/N\gtrsim25$ per pixel and measure stellar parameters with a precision of $\sigma (\Teff , \log g , \textrm{[Fe/H]}) \la (\SI{100}{\kelvin}, \SI{0.2}{dex}, \SI{0.1}{dex})$.

We start the preparation of these spectra with the same normalisation process described above (\Sref{sec:Ref_spec}). The only difference in application is that we need wider masked wavelength regions for hydrogen lines and telluric absorption because the radial velocity of these spectra is unknown at this point of the analysis.
We adjusted each mask to accommodate a $\SI{250}{\kilo\metre\per\second}$ maximum radial velocity (and barycentric velocity) based on expected radial velocities of stars closer to the Galactic Centre. These larger masks potentially cause issues for the continuum fit of individual sub-spectra but they can be avoided with correct sub-spectrum positioning, i.e.\ hydrogen lines and telluric features should not be on the edge of sub-spectra. Note that we apply a more precise local normalisation around each stellar absorption feature (\Sref{sec:Precise_renorm}) so this initial normalisation does not need to be perfect.
Furthermore, we identified sky emission features that tend not to be well corrected and interfere with the subsequent preparation process in \Sref{sec:measurement_EW}. We normalise a non-sky subtracted version of the target spectrum which makes identification of emission features trivial and remove pixels within $\SI{30}{\kilo\metre\per\second}$ of sky emission from all further analysis.

We need to correct the radial velocity of a target spectrum to the zero-velocity frame of the reference spectrum.
We selected a $\SI{1600}{\kilo\metre\per\second}$ velocity range in each band of HERMES centred on these four wavelengths: $4807, 5763, 6670 \textrm{ and } \SI{7818}{\angstrom}$. These were chosen because they are surrounded by a relatively large number of stellar absorption features, which is ideal for a cross-correlation.
We cut out the corresponding wavelength range in the reference and target spectra and reduce the resolving power of the reference sub-spectrum to match the target spectrum (\Sref{sec:res_pow_align} for details).
The $\SI{1600}{\kilo\metre\per\second}$ sections of the reference and target spectra are then cross-correlated to find the radial velocity shift between them.
We apply the correction as a constant shift in wavelength to the whole band to keep a constant pixel wavelength spacing, as opposed to a more precise wavelength dependent radial velocity correction.
This makes it only precise to $\sim 1-2$ HERMES pixels, which is sufficiently accurate for our purposes as we later apply an additional line-by-line correction for the radial velocity (\Sref{sec:rv_precise}).

\subsubsection{Library/stacked spectra}
The stacked spectra of \citet[][]{Zwitter2018}, which are used as a spectral library to create the model in \Sref{sec:stellar_par}, need to undergo the same normalisation and radial velocity correction steps as the target HERMES spectra (\Sref{sec:tar_prep}).
Additionally, we need to create an error array for these spectra so that we can later create a realistic pixel weight array for individual absorption features (\Sref{sec:EW_measure}). Any error array proportional to $\sqrt{\textrm{flux}}$ simulates a real error array well enough for this purpose.
Furthermore, we re-sample the stacked spectra to a grid with 4096 pixels per band and constant pixel width using SpectRes to closely match the properties of HERMES spectra.

\subsubsection{High resolution target spectra}\label{sec:high_res_prep}
High resolution target spectra are primarily used for testing purposes in EPIC. The spectral data can be obtained from any instrument with higher resolving power than HERMES ($R>28{,}000$). We primarily use spectra from ESO's (European Southern Observatory) HARPS spectrograph (High Accuracy Radial velocity Planet Searcher) as the data are easily accessible from the ESO Science Archive and the instrument has accumulated a large number of high $S/N$ spectra over almost 20 years of operation.
We only use spectra with $S/N\ga 50$ per $\SI{0.8}{\kilo\metre\per\second}$ pixel, which further increases by a factor of $\sim2$ when they are convolved and down-sampled to match HERMES resolution and sampling, so we can safely assume these spectra to be noiseless for the purpose of our tests.

The preparation of these spectra involves steps already explained in \Sref{sec:Ref_spec} and \ref{sec:tar_prep}, i.e.\ converting the wavelength range to HERMES wavelength bands, normalising the spectrum and correcting for radial velocity. Note that the infrared wavelength region is not covered by HARPS, so we are left with only the blue, green and red bands.
In contrast to \Sref{sec:Ref_spec}, these are intended to be used as target spectra, so we have to reduce their resolving power to match that of HERMES. So we convolve each band with a Gaussian kernel that reduces the resolving power to $28{,}000$. The convolution process is detailed in \Sref{sec:res_pow_align}.

\subsection{Selection of absorption features}\label{sec:linelist}
The list of stellar absorption features (`line list') used in this work is specialised for solar analogue spectra from the HERMES spectrograph. We included lines from two previous line lists: the first was used by \citet{Datson2015} specifically to help identify solar analogue stars --- 82 absorption lines from this list were available within the HERMES wavelength range; the second line list was used by the GALAH survey \citep[][]{Buder2018} and provides 216 absorption lines that are useful to determine stellar atmospheric parameters. Note that 44 features were present in both lists.

Additionally, we visually selected other features within the HERMES wavelength bands that were not included in either of these lists, which contributed another 54 absorption features for a total of 308.
Of these lines, we selected only those matching the following criteria, each applied to the features as they appear in the solar reference spectrum at HERMES resolution (unless otherwise specified):
\begin{enumerate}[labelwidth=*, leftmargin=*]
    \item We check if the feature is useful in a low $S/N$ ($\sim 25$) HERMES spectrum. Our measurements indicate that the typical uncertainty for EWs in such a spectrum is $\sigma_\textrm{EW}\sim$1\,m\AA. We calculate the EW of each line and remove those with EW$\leq 5$\,m\AA\, from the list.
    \item The features should not be saturated, so we visually determine if a line is saturated or close to saturation and remove them from the list. For this selection we make use of the high resolution ($R = 200{,}000$--$300{,}000$) solar atlas \citep[][]{Chance2010}. We defined saturation in a line as absorbing more than 80\% of the normalised flux at its centre.
    \item We reject features that appear to be blended with other features. This is done by visual inspection as it does not need to be rigorously defined. Blends can lead to more complicated behaviour with respect to stellar parameters, so we also select against them in the residual based selection below.
    \item We avoid features that are close ($\sim \SI{50}{\kilo\metre\per\second}$) to telluric absorption to avoid atmospheric influences.
\end{enumerate}
A feature's EW also needs to show simple behaviour with respect to changing stellar parameters so that we can establish a model for it -- see \Sref{sec:stellar_par}. We inspected residuals between the model and test data and determined that lines with average residuals of more than 2\% of their full EW range are filtered out. This leaves us with 125 absorption features in the line list.

We chose to apply the above criteria less strictly to ionised lines -- i.e. average residuals can have up to 3\% of the line's full EW range. This is done because ionised lines are more sensitive to the surface gravity than neutral lines, so our aim was to increase the method's overall sensitivity to differences in $\log g$ between stars. This means that we extended the line list by an additional 6 ionised lines (from 125 to 131 total lines). When comparing the application of the EPIC algorithm with and without these additional lines the resulting uncertainties for all stellar parameters are reduced: $\sigma \Teff$ by 2\%, $\sigma\textrm{[Fe/H]}$ by 2.5\% and $\sigma\log g$ by 7\% with the additional ionised lines.
Our final line list of 131 features is provided in \Tref{tab:linelist}.

\subsection{Line preparation}\label{sec:measurement_EW}
In this section, we describe how the wavelength area around each spectral feature is prepared so that EW measurements can be made consistently in the library/stacked and target spectra.
Each line is analysed independently in each spectrum.

\begin{figure}
    \centering
    \includegraphics[width=0.43\textwidth]{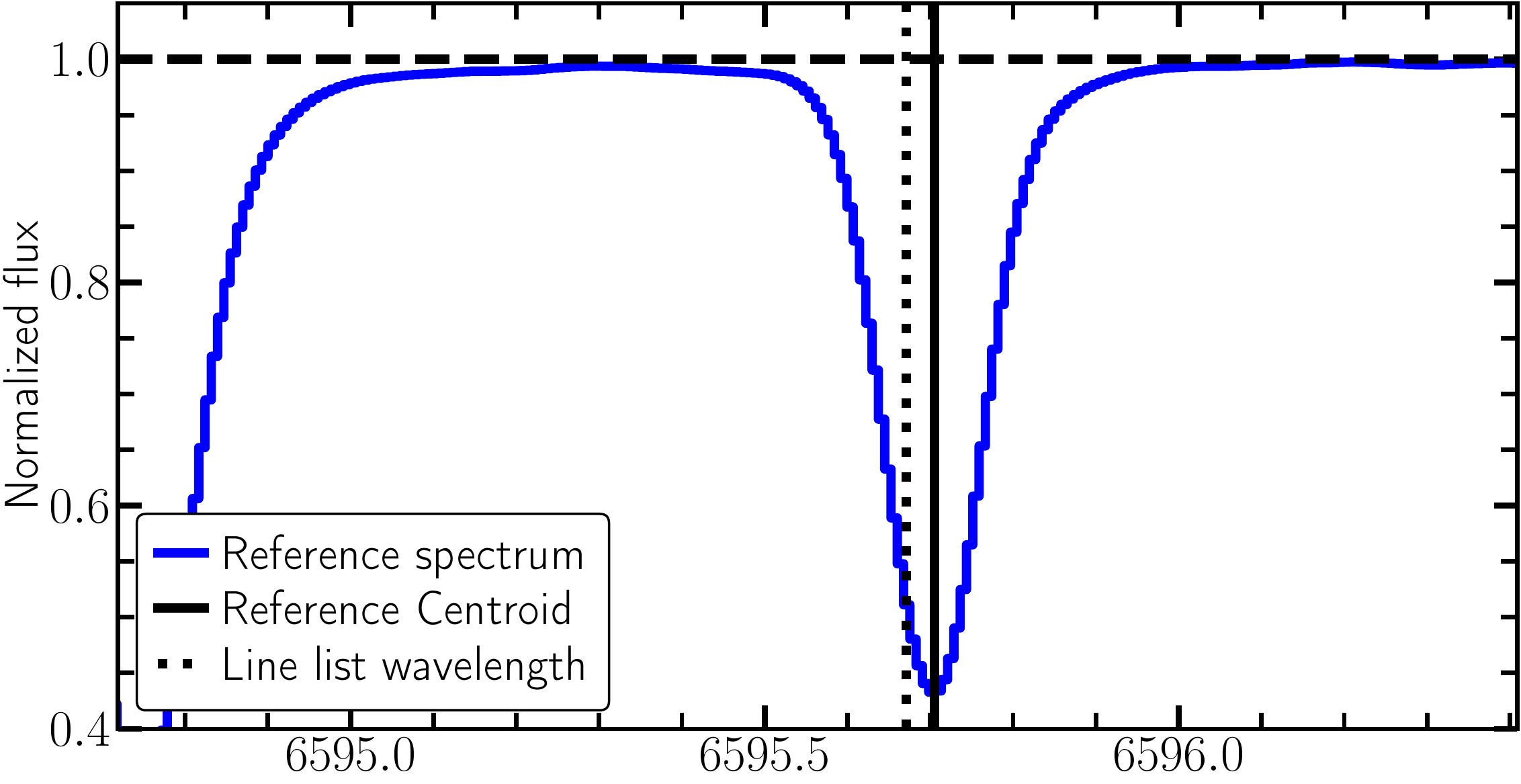}
    \includegraphics[width=0.43\textwidth]{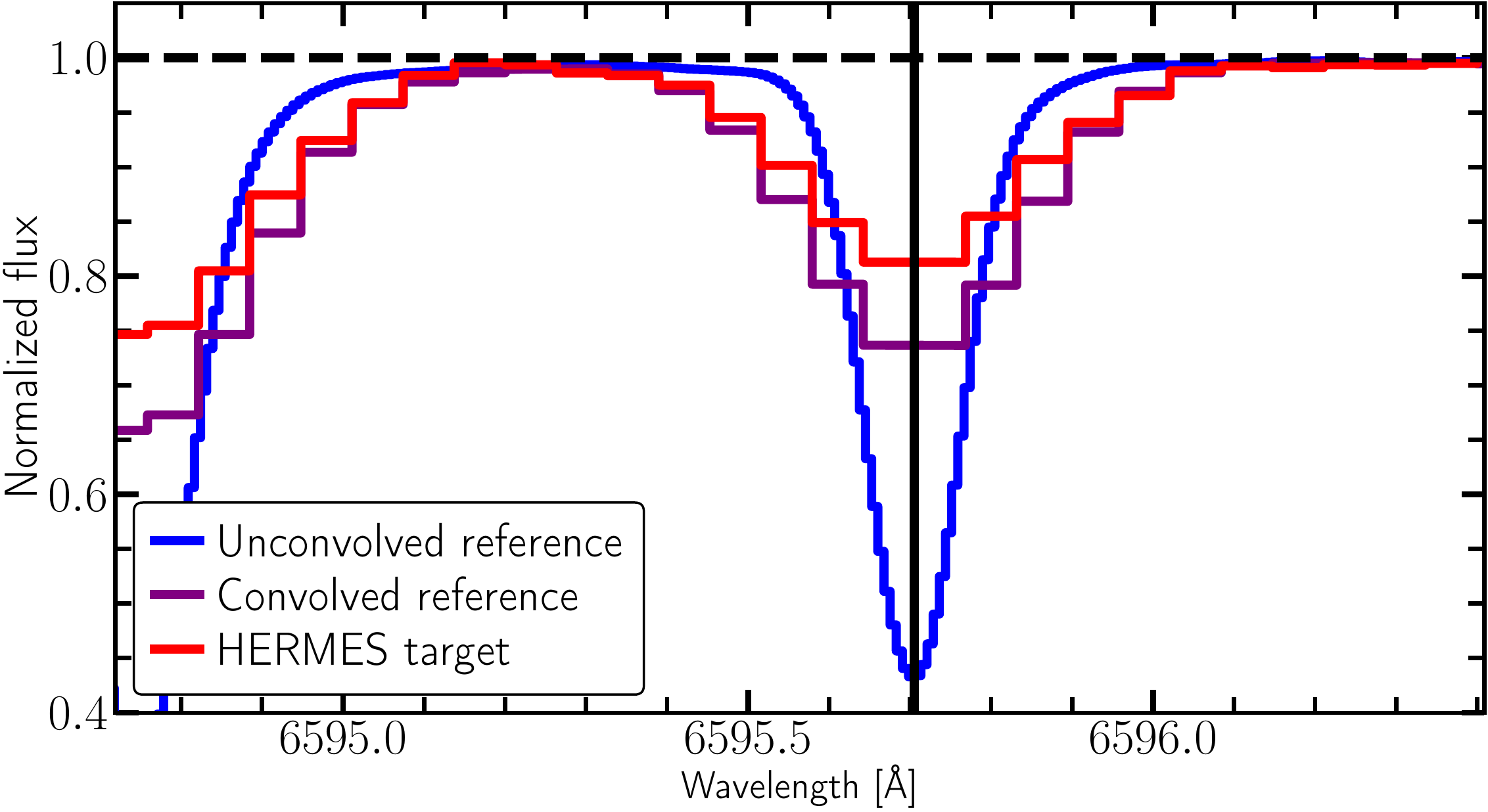}
    \includegraphics[width=0.43\textwidth]{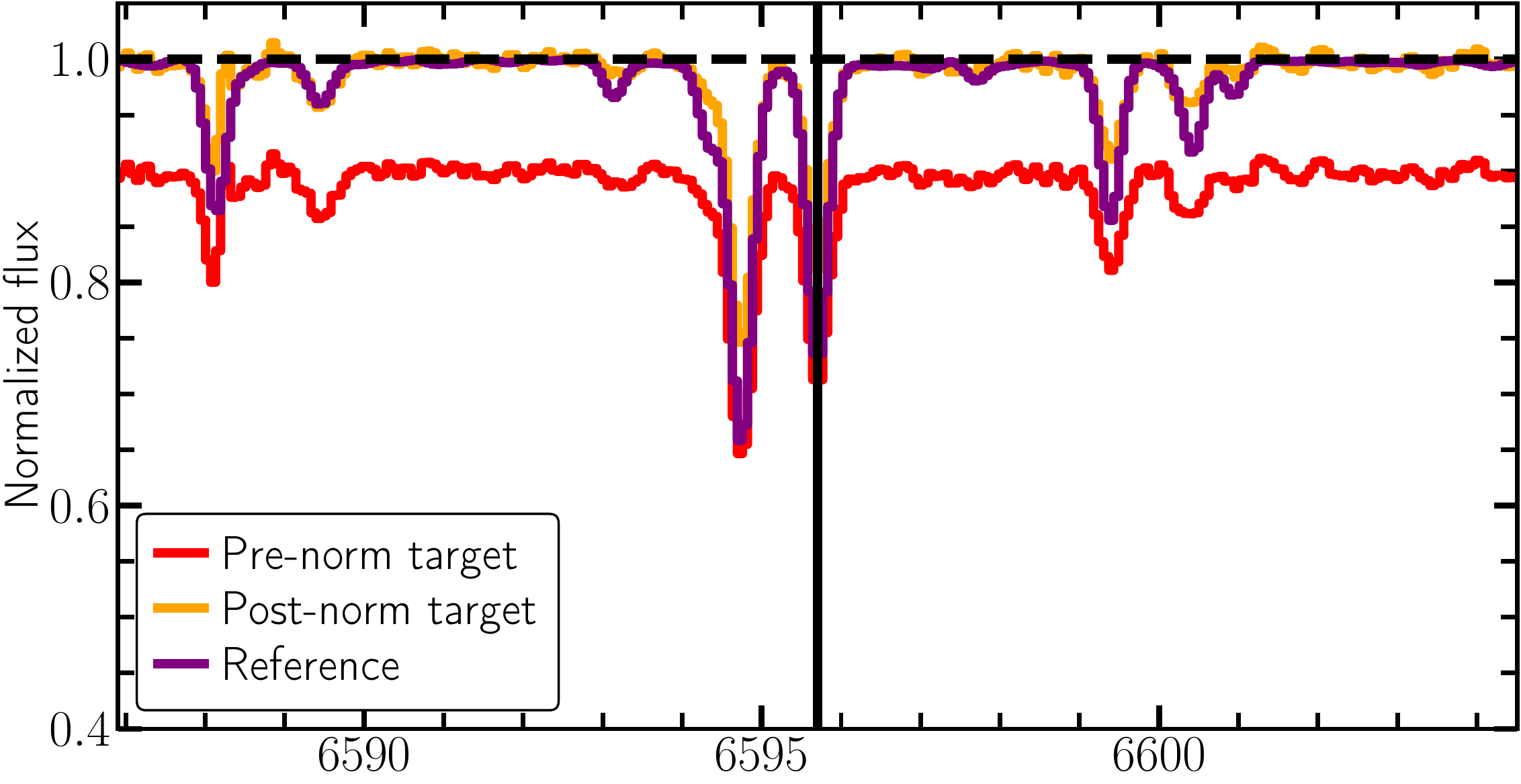}
    \includegraphics[width=0.43\textwidth]{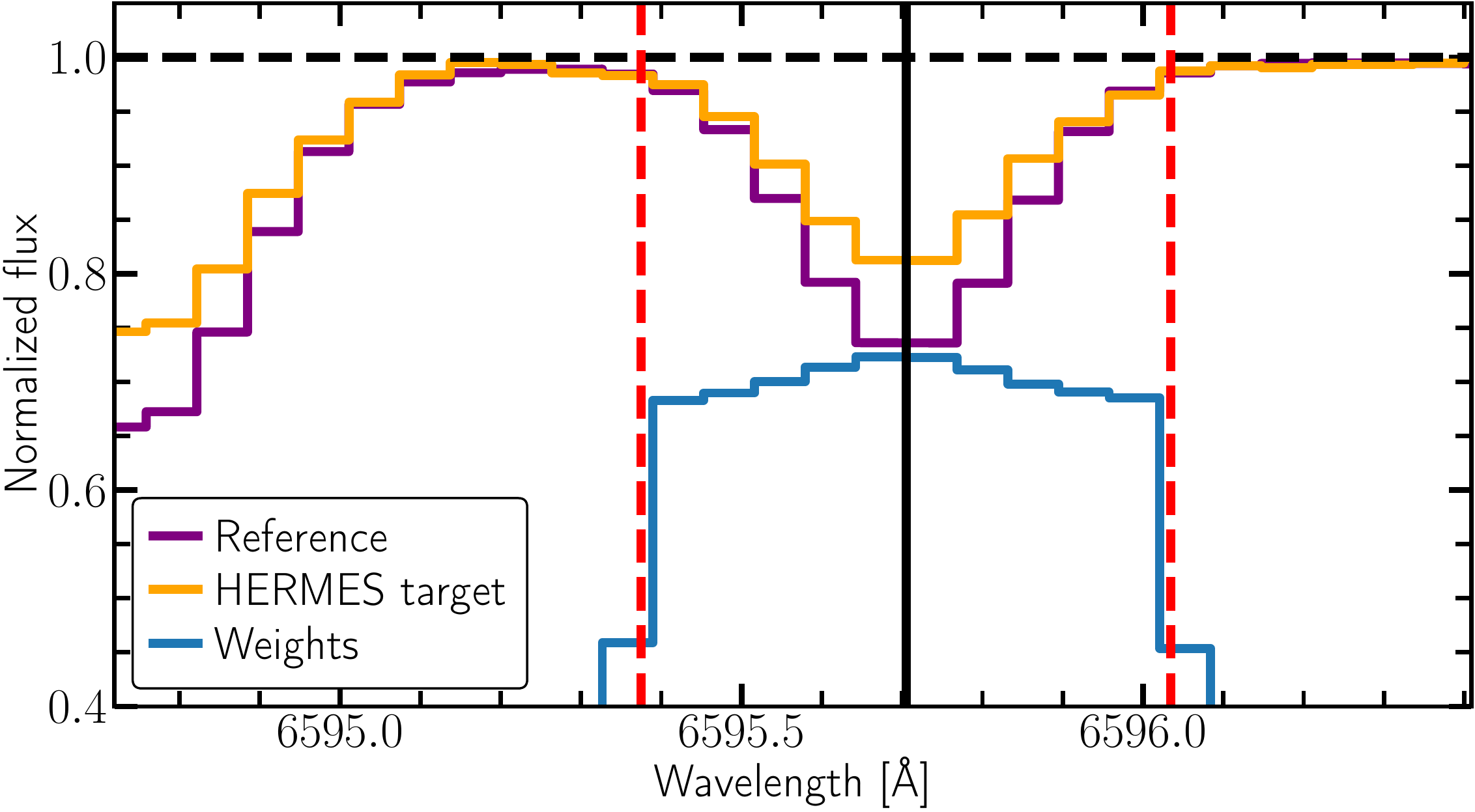}
    \caption{Line preparation and EW measurement in EPIC. The top panel shows the line centering. The algorithm searches for the lowest flux value in the local environment of the reference spectrum with sub-pixel precision (\Sref{sec:line_centre}). The black dotted vertical line shows the line list information and the solid black line is our newly determined line centroid.
    The second panel visualises the resolving power matching and wavelength grid re-sampling of the reference solar spectrum to the HERMES target spectrum (\Sref{sec:res_pow_align}).
    The third panel shows the re-normalisation process for the target spectrum, which matches the continuum flux with the reference (effect exaggerated for visual purposes; \Sref{sec:Precise_renorm}).
    The bottom panel shows the EW window (red dashed lines) and weighting of each pixel for the EW measurement (blue line, \Sref{sec:EW_measure}).}
    \label{fig:line_calibration}
\end{figure}

\subsubsection{Precise radial velocity correction} \label{sec:rv_precise}
We determined the radial velocity using a $1{,}\SI{600}{\kilo\metre\per\second}$ window in each band as part of the spectral preparation (\Sref{sec:tar_prep}), but we need to ensure that each absorption feature in the reference and target spectrum is aligned to minimise EW systematic errors.
In this step we make a more precise local radial velocity correction for each line of interest.
For a comparatively small velocity window of $\SI{380}{\kilo\metre\per\second}$ around the line, we re-sample the target or stacked spectrum using SpectRes to match the wavelength grid of the reference spectrum. This allows us to gain sub-HERMES pixel precision in the new correction.
We apply the same cross-correlation process from \Sref{sec:tar_prep} to this small wavelength section of re-sampled target or stacked spectrum to calculate the radial velocity differences between the reference and target absorption feature, which is then corrected for.

\subsubsection{Line centering}\label{sec:line_centre}
We need to determine the precise absorption line centre so that we can define the wavelength window for the EW measurement. We use the wavelength values in the line list as a starting point for this determination. The top panel of \Fref{fig:line_calibration} shows an example line for which we determined the line centre.

The absorption features of reference and target spectrum are aligned with each other after the previous step so we only need to determine the line centre for one of them. We choose to determine the line centroid in the reference spectrum to avoid noise. We define a small wavelength window of $\SI{16}{\kilo\metre\per\second}$ around the line list wavelength for the feature. This window is chosen as a little more than half the full width at half maximum (FWHM) at HERMES resolution and assumes that the line list wavelength is within the absorption feature.
We identify the line centroid as the minimum of a quadratic fit to the three pixels with the lowest flux value within the chosen window.

\subsubsection{Resolving power reduction and reference re-sampling}\label{sec:res_pow_align}
We reduce the resolving power $R$ of the reference spectrum to match the target spectrum, as visualised in the second panel of \Fref{fig:line_calibration}.
As noted in \Sref{sec:Ref_spec} the reference spectrum has a high resolving power (i.e.\ $R = 200{,}000$--$300{,}000$) so we convolve it with a Gaussian with a FWHM corresponding to HERMES spectra. We adjust the FWHM to take the finite $R$ of the reference spectrum into account. Importantly, HERMES does not have a constant resolving power as a function of wavelength \citep[][]{Kos2017}, so we need to determine the resolving power for each line separately. We do this by using the current version of the GALAH resolving power maps which provide $R$ for all HERMES fibres as functions of wavelength \citep[see section 7 in][]{Kos2017}. 
Note that the original, observed wavelength was used to find the corresponding $R$ value in the map, not the radial velocity corrected value.

Furthermore, we need the reference spectrum wavelength grid to match that of the target spectrum, so after the convolution is applied we use SpectRes to re-sample the reference spectrum to the target or stacked spectrum's wavelength grid.

\subsubsection{Continuum re-normalisation}\label{sec:Precise_renorm}
We re-normalise the target/stacked spectrum locally to ensure that its continuum matches the reference spectrum's as closely as possible. This improves upon the initial continuum in the spectral preparation (\Sref{sec:Ref_spec}). The third panel of 
\Fref{fig:line_calibration} visualises the effect of this step.

We choose a wavelength area around the absorption line with a width of $\SI{800}{\kilo\metre\per\second}$. This window is chosen to be large because it allows a noise-resistant, precise matching of the target continuum to the reference one. However, it is limited to this size so that it is not sensitive to poor initial continuum fits. Additionally, this wavelength space determines how close to the edge of a spectral band our absorption features can be, i.e.\ not closer than $\SI{800}{\kilo\metre\per\second}$ as to not affect this process.

For convenience, we first re-normalise the reference spectrum by the maximum flux in the chosen window.
The high $S/N$ reference spectrum is not affected by any cosmic rays and has a precise sky subtraction of telluric emission, so this straight forward approach is sufficient.

Next, EPIC adjusts the average target spectrum flux to that of the re-normalised version of the reference in the same wavelength region around the absorption line (i.e.\ $\SI{800}{\kilo\metre\per\second}$).
We remove the 75\% of pixels with the lowest flux values in the reference spectrum from both the target and the reference in order to ignore absorption lines in this process (note that their wavelength grids are identical at this point in the analysis). Additionally, we remove the $\SI{30}{\kilo\metre\per\second}$ window around the line centroid used for the EW measurement (\Sref{sec:EW_measure}).
A weight is assigned to all remaining pixels, given by their inverse flux variance in the target spectrum. We determine the weighted mean flux of all remaining pixels in both the reference and target spectrum. The target spectrum is then scaled by the ratio of these weighted means.
This method of normalisation aligns the continuum of the target spectrum very precisely with the reference around the desired absorption line.

\subsection{Equivalent width measurement}\label{sec:EW_measure}
After the preparation steps in \Sref{sec:preparing} and \ref{sec:measurement_EW}, we can define a window of $\SI{30}{\kilo\metre\per\second}$ around the line centroid in which to measure the EW. 
This window is optimised to include most of the absorption of the feature while minimising the influence of potential weak blending effects (lines with strong blending were excluded in \Sref{sec:linelist}).

We determine a weight $w_{i}$ for each pixel,
\begin{gather}
    w_{i} = \frac{b_i}{\sigma_i^2},
\end{gather}
where $\sigma_i$ is the uncertainty of the flux and $b_i$ is the fraction of the pixel-width within the defined wavelength window. That is, pixels fully within the line definition have $b_i=1$, pixels fully outside have $b_i=0$ and pixels partially within have $0<b_i<1$. A typical weight distribution is visualised in the bottom panel of \Fref{fig:line_calibration}.
These weights are derived with the flux uncertainties in the target spectrum but applied to both the target and reference spectrum to maximise the comparability between the two.

With a weight attached to every pixel, we can calculate the normalised absorption and the EW of the line via
\begin{gather}
	A = 1 - \left(\frac{\sum\limits_i f_i  w_i}{\sum\limits_i w_i}\right), \\
	EW = A \: W,
\end{gather}
where $A$ is the normalised and weighted average absorption, $W$ the width of the defined wavelength window and $f_i$ the flux in an individual pixel normalised to 1.
For the selected absorption features, we use the difference between their EW and that in the reference spectrum, $\delta$EW, in further analysis to measure the stellar parameters.

\subsection{Measuring stellar parameters}\label{sec:stellar_par}
\begin{figure}
    \centering
    \includegraphics[width=0.45\textwidth]{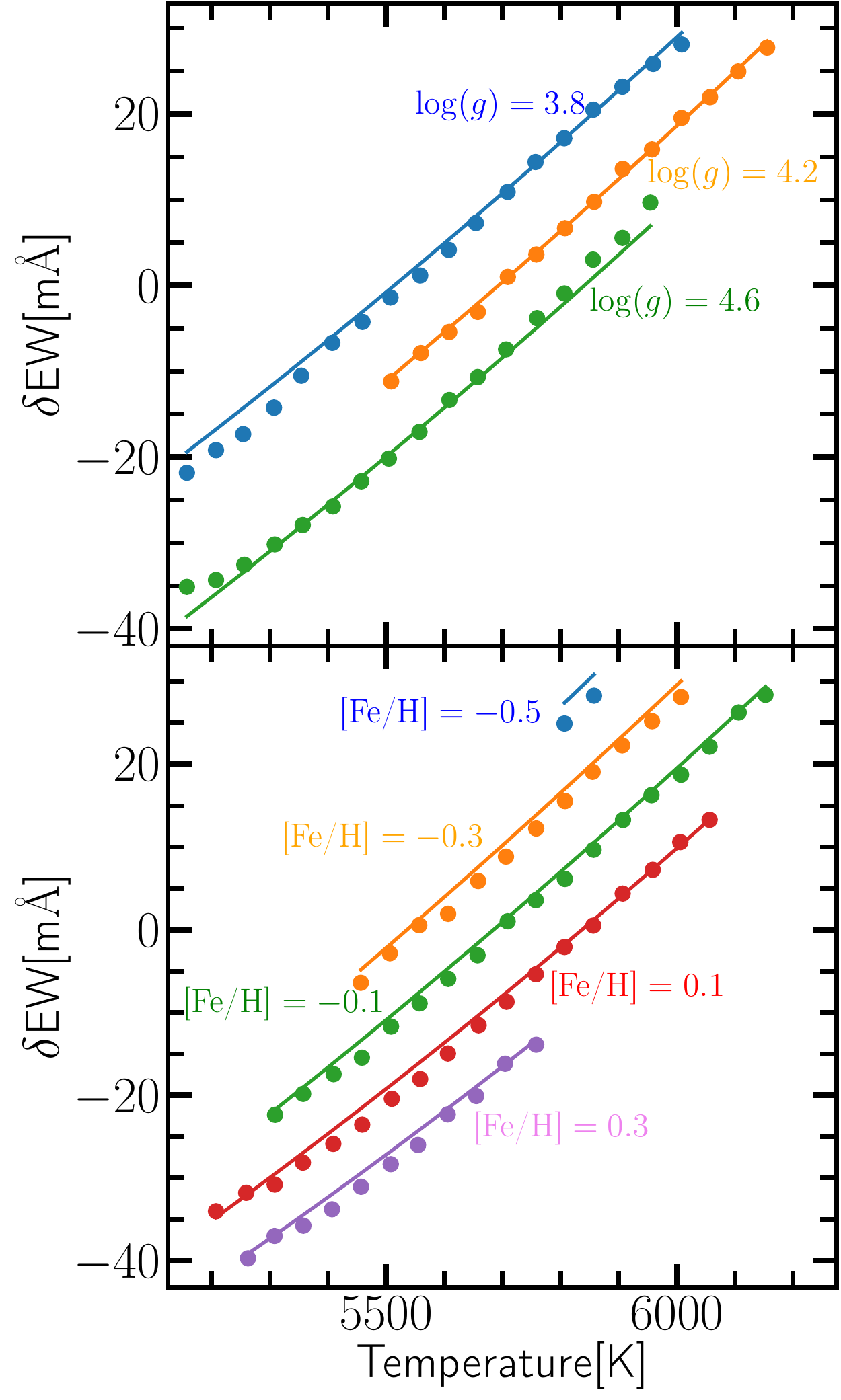}
    \caption{Stellar parameter model for $\Teff$ of a single transition (Fe I at $\SI{4892.86}{\angstrom}$). The plot shows slices of a four dimensional fit of the three stellar atmospheric parameters to the EW differences (between the stacked spectra and the solar reference spectrum). The top panel has constant $\textrm{[Fe/H]} = \SI{0.0}{dex}$ while the bottom panel has constant $\log g=\SI{4.4}{dex}$. The solid lines represent the model fit while the dots are EW measurements for stacked spectra from \citet[][]{Zwitter2018}. The colours correspond to constant values in $\log g$ (top panel) and [Fe/H] (bottom panel) as denoted close to each line.}
    \label{fig:model_show}
\end{figure}

Each stellar absorption line has a unique dependency on the stellar parameters. The EPIC algorithm models these dependencies and uses them to measure stellar parameters via EWs.

First, we require a set of spectra with known stellar parameters in order to determine the coefficients of the model (see \Eref{eq:model1}). The GALAH survey provided medians of observed spectra \citep{Zwitter2018} with a wide range of stellar parameters; ideal for the purpose of model creation.
These spectra were made from spectra in GALAH DR2 \citep{Buder2018}. The stellar parameters of the initial spectra have been determined by The Cannon \citep{cannon2015} and sorted into bins by their effective temperature, surface gravity and metallicity \citep[][]{Zwitter2018}. The size of these bins in stellar parameter space is $\Teff=\SI{50}{\kelvin}$, $\log g=\SI{0.2}{dex}$ (g in cm/s$^2$) and $\textrm{[Fe/H]}=\SI{0.1}{dex}$. All spectra within the same bin were stacked to create high $S/N$ model spectra with known average stellar parameters.
We use the weighted mean stellar parameters for spectra within a bin instead of the central value of the bin in our calculations. This is necessary as stellar parameters are not uniformly distributed within a bin, e.g.\ more spectra within GALAH DR2 have $\log g=4.3 \textrm{ -- } \SI{4.4}{dex}$ than $\log g=4.4 \textrm{ -- } \SI{4.5}{dex}$ meaning a stacked spectrum combined from all those spectra will have an effective surface gravity $\log g<\SI{4.4}{dex}$ while the bin is labelled as $\log g=\SI{4.4}{dex}$. While we have chosen to use the weighted mean values here, we also conducted the calibration process using the median instead, and found that the resulting stellar parameters for target stars differ by $\delta(\Teff , \log g, \textrm{[Fe/H]}) \lesssim (\SI{2}{\kelvin}, \SI{0.005}{dex}, \SI{0.001}{dex})$, which are negligible compared to our other sources of error.

With these spectra we can create a model for each individual feature within the line list. Each model has different parameters ($c_i$), which are used to connect effective temperature $\Teff$, surface gravity $\log g$ and metallicity [Fe/H] to varying $\delta$EW. We attempt to maximise the model's ability to reproduce the training set while using the simplest formula for a fitting approach:
\begin{align} \label{eq:model1}
\delta EW &= c_0 + c_1 \Teff + c_2 T^2_\textrm{eff} + c_3 \log g + c_4 \log g^2
 \notag \\ &+ c_5 [\textrm{Fe/H}] + c_6 [\textrm{Fe/H}]^2 + c_7 \frac{[\textrm{Fe/H}]}{\Teff}
\end{align} 
where $c_{i}$ are the model constants for an individual line. We visualise the model creation in \Fref{fig:model_show} which shows a single transition for which we measured EWs in stacked spectra.
We tried several different model formulas to fit the measurements and found that second order moments and a cross term between [Fe/H] and \Teff\ improved our results. Additional terms, including cross-terms involving $\log g$, had a negligible effect on our results and added instability to the method.

We determined the EWs for all absorption features within our line list for all available stacked spectra that have stellar parameters within a initially broad range around solar values, i.e.\ $\Teff = \SI{5800}{\kelvin} \pm \SI{700}{\kelvin}, \log g = \SI{4.4}{dex} \pm \SI{0.6}{dex} \textrm{ and [Fe/H]}= \SI{0.0}{dex} \pm \SI{0.5}{dex}$. It is possible to extend the model boundaries (e.g.\ to $\Teff = \SI{5800}{\kelvin} \pm \SI{1400}{\kelvin}$), but we would need to use a more complex model than \Eref{eq:model1} to avoid systematic errors and more available stacked spectra to cover that range. 
We can solve \Eref{eq:model1} for all $c_{i}$ to connect EWs to stellar parameters for each line. This is done via a fit for these parameters that is based on the measurement of EWs and the stellar parameters that are already known for the stacked spectra.
With these model parameters we can calculate stellar parameters for stars by reversing the procedure and solving for $\Teff$, $\log g$ and [Fe/H] in \Eref{eq:model1}. When reversing the process, we perform a least squares fit to the $\delta$EW values of all lines in a target spectrum, with inverse variance weighting which combines the uncertainties in the EW measurements and model values. The least squares fitting provides the best-fit stellar parameters and the covariance matrix, the diagonal terms of which provide our quoted uncertainties.

In principle, it is possible to extend the number of stellar parameters that are measured by the model (e.g. by specific metal abundances), but there are several points against this. First, the uncertainty would increase for all stellar parameters and $c_{i}$ parameters. Furthermore, we would need a complete set of spectra which have these parameters calculated as a base for further analysis. So it is not practical at the current time to expand the algorithm this way.

\subsection{Stellar parameter calibration}\label{sec:calibration}
The stellar parameters calculated with EPIC inherit any potential systematic errors from the stacked spectra \citep{Zwitter2018} and their calibration from The Cannon \citep[][]{cannon2015}. Therefore, we need to investigate and correct systematic effects if present.
First, we found differences between EPIC's stellar parameters measured for solar spectra and their literature values for the Sun. We implemented a `reference-point correction' for this (see \Sref{sec:solar_test} for details).
Additionally, we investigated stellar parameter dependent differences between EPIC's estimated values and stellar parameters measured from high resolution and high $S/N$ echelle spectroscopy \citep[][]{Casali2020}. In addition to the `reference-point correction' we perform a `higher-order correction' to account for stellar parameter dependent systematic errors (see \Sref{sec:High_precision} for details). This enables us to switch from an initial calibration based on the spectral library to one that agrees more with stellar parameters determined in \citet[][]{Casali2020}.

For all of these calibrations we use spectra with simulated error arrays using values corresponding to S/N = 50 per pixel in all HERMES bands. This provides a reasonably realistic relative weighting of the flux and model errors in the EWs when fitting the model. Using a very large $S/N$ would have left the model errors to dominate. So, instead, we chose a $S/N$ that is more typical of the real HERMES spectra we will normally apply EPIC to. The stellar parameters change very little as a function of this chosen S/N in the range 20--100 per pixel: $\delta(\Teff, \log g$, [Fe/H]$) = (\SI{3}{\kelvin}, \SI{0.002}{dex}, \SI{0.0008}{dex})$. If spectra of higher $S/N$ are to be analysed, EPIC would need re-calibration or, in future versions, possibly a $S/N$-dependent calibration.

\section{Results}\label{sec:Results}
We applied EPIC to spectral data to test and refine the calibration, and present the full capability of the method. We describe the results of the algorithm for the following data products: the solar reference spectrum \citep[][]{Chance2010} and solar spectra obtained from collecting reflected light from minor solar system bodies -- Ganymede, Ceres and the Moon -- from HARPS \citep[][]{Mayor2003}, spectra from HARPS for which precise stellar parameters have been measured \citep[][]{Casali2020} and HERMES spectra from the GALAH Data Release 2 catalogue \citep[][]{Buder2018}.

\subsection{Solar spectra: reference-point correction}\label{sec:solar_test}
\begin{figure}
	\centering
	\includegraphics[width=0.45\textwidth]{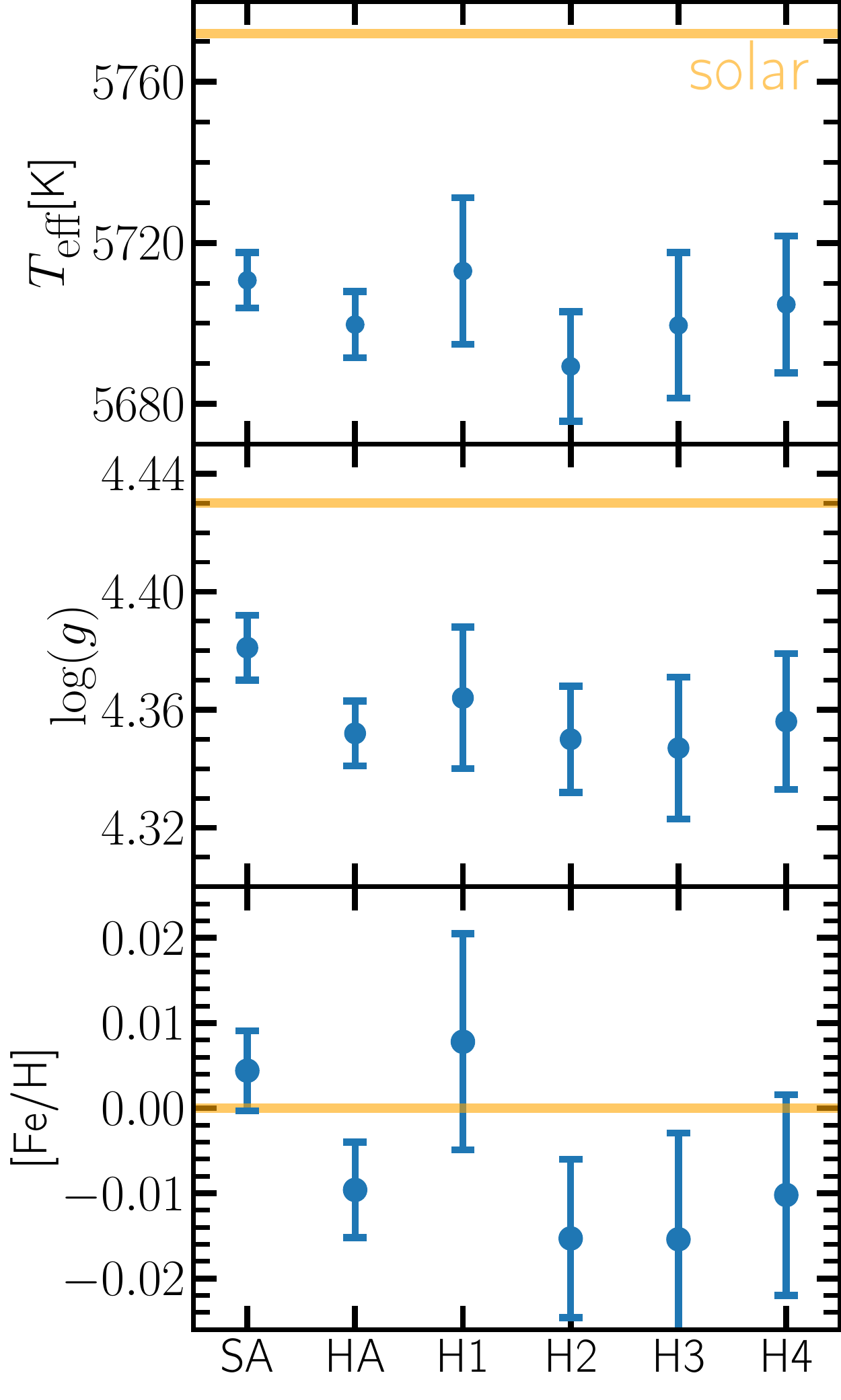}
	\caption{EPIC uncalibrated stellar parameter measurements for solar spectra. This figure shows results for the following solar spectra, each analysed with the HARPS version of our line list (i.e.\ excluding absorption features marked with an asterisk in \Tref{tab:linelist}): SA: solar atlas \protect\citep{Chance2010}; H1-4: HARPS spectra obtained by observing reflected light from minor bodies satellites (H1: Ceres observed in 2006; H2: Ganymede in 2007; H3: the Moon in 2008; H4: Ceres in 2009); HA: mean value of H1-4 weighed by their inverse variance.}
	\label{fig:solar}
\end{figure}

The main function of EPIC is to measure stellar parameters, so it is important to test if measured parameters match their literature values for selected test targets. 
Naturally, the first test is to reproduce stellar parameters of the Sun as it is the main reference for our method.
We apply EPIC to the solar atlas used as our main reference \citep{Chance2010} which is prepared as a reference (\Sref{sec:Ref_spec}) and a target (\Sref{sec:high_res_prep}) spectrum for this analysis.
We derived stellar parameters using the EPIC algorithm and find that the values calculated for the solar atlas ($\Teff, \log g$, [Fe/H]$ = 5711 \pm \SI{7}{\kelvin}, 4.38 \pm \SI{0.01}{dex}, 0.004 \pm \SI{0.005}{dex}$) are offset from the current IAU standard values for solar stellar parameters $(\Teffnom, \log g_\odot$, [Fe/H]$_\odot) = (\SI{5772}{\kelvin}, \SI{4.44}{dex}, \SI{0.0}{dex})$ in \citet[][]{Prsa2016} by $\Delta (\Teff, \log g$, [Fe/H]$) = (\SI{61}{\kelvin}, \SI{0.06}{dex}, -\SI{0.004}{dex})$.
We are able to trace these offsets back to the initial stellar parameter measurements of The Cannon which show systematic differences when compared to other stellar parameter studies \citep[e.g.][more in \Sref{sec:High_precision} and \Sref{sec:Cannon}]{Adibekyan2017, Casali2020}.
Therefore, we need to address this issue with a `reference-point correction'. This process corrects the model (\Sref{sec:stellar_par}) by adding constants to the stellar parameters of model spectra, i.e.\ we changed $\Teff \rightarrow \Teff + \SI{61}{\kelvin}, \log g \rightarrow \log g + \SI{0.06}{dex} \textrm{ and [Fe/H]} \rightarrow \textrm{[Fe/H]} - \SI{0.004}{dex}$ in \Eref{eq:model1}. This corrects the measured solar stellar parameters to their literature values.
We note that \citet{Buder2018} found the same mean offset (61\,K) between the spectroscopic $T_{\rm eff}$ in GALAH (DR2) and that derived via the infrared flux method \citep{Casagrande2010, Casagrande2014} using SkyMapper photometry.

As a second test, we compare the stellar parameters measured for the solar atlas to those we measured for another set of solar spectra to verify the `reference-point correction'.
We utilise the archive of the HARPS spectrograph, i.e.\ their reflected solar spectra from Ceres, Ganymede and the Moon. These are high resolution spectra with high $S/N$ that can easily be convolved to HERMES resolution (\Sref{sec:high_res_prep}).
A shortcoming of these spectra is that HARPS only covers the wavelength ranges of the B, G and R bands in HERMES, so the IR band is not available.
Another shortcoming is that the data pipeline for HARPS does not provide a correction for telluric absorption. Therefore, we need to adjust the line list, removing features that can fall within the same wavelength range as telluric absorption due to stellar radial velocity differences and Earth's barycentric velocity variations. We considered a maximum radial velocity difference of $\SI{200}{\kilo\metre\per\second}$ for this alternative line list in order to accommodate the HARPS spectra from \citet[][]{Casali2020} considered in \Sref{sec:High_precision}. All features that we excluded for the analysis with HARPS spectra are marked with an asterisk in \Tref{tab:linelist}.
The purpose of this test is to compare measured stellar parameters for the solar atlas \citep[][]{Chance2010} to those of the HARPS solar spectra, so we need to apply the same analysis to both, i.e.\ we only analyse absorption features for the solar atlas that can be used for HARPS spectra. 

The results of this analysis can be seen in \Fref{fig:solar}.
We identify a discrepancy, i.e.\ $\Delta (\Teff, \log g$,[Fe/H]$) = \SI{-11}{\kelvin}, \SI{-0.029}{dex}, \SI{-0.0140}{dex}$, between stellar parameters measured for HARPS spectra and the solar atlas (HA and SA in \Fref{fig:solar}). We traced these differences back to slight changes in the normalisation process (\Sref{sec:Precise_renorm}) that are caused by very weak telluric features that have not been removed in HARPS spectra.
Additionally, when analysing the solar atlas with and without an IR band, i.e.\ with the reduced line list, we found negligible stellar parameter differences of $\Delta (\Teff, \log g$,[Fe/H]$) = \SI{-0.6}{\kelvin}, \SI{-0.0003}{dex}, \SI{-0.0024}{dex}$. However, a missing IR band leads to an increase of $\sim20\%$ in uncertainties which is expected as the number of absorption features declines. The same can be observed for other HERMES spectra.

\subsection{Systematic effects between HERMES and HARPS}\label{sec:HAR_HER_test}
\begin{figure}
	\centering
	\includegraphics[width=0.475\textwidth]{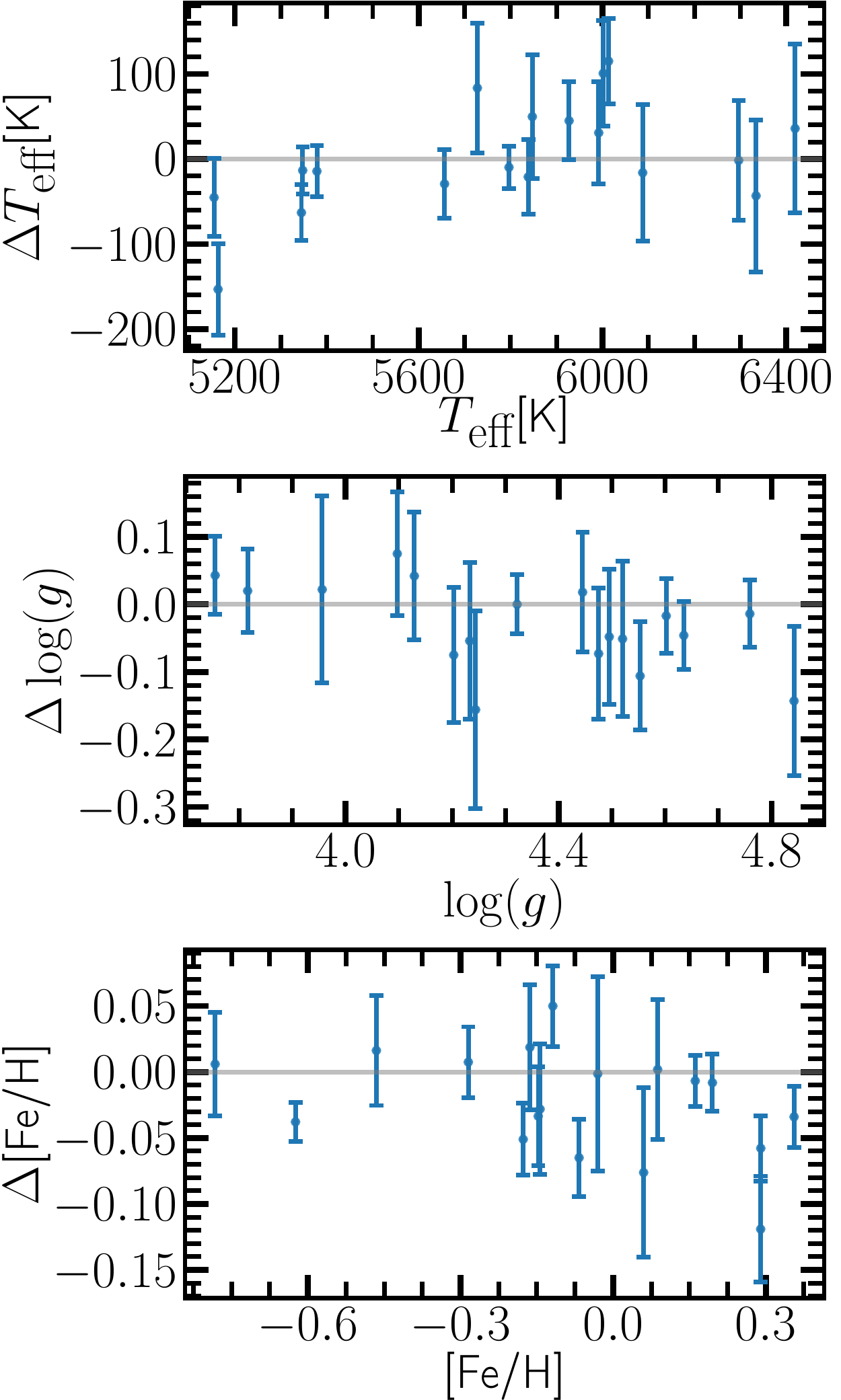}
	\caption{Differences between EPIC results when using HERMES spectra compared to HARPS spectra. The three panels show the effective temperature, surface gravity (g in cm/s$^2$) and metallicity. The y-axis shows the difference of these parameters between HARPS and HERMES spectra for the same target stars. The y-error-bars are combined from both the HERMES and HARPS uncertainties. The x-axis is the respective stellar parameter as measured for the HARPS spectra with EPIC. The `reference-point correction' has been applied prior to this analysis while the `higher-order correction' (\Sref{sec:High_precision}) has yet to be applied.}
	\label{fig:harps_hermes}
\end{figure}
We used spectral data from the HARPS spectrograph in the process of testing EPIC (\Sref{sec:high_res_prep}) which raises the question of whether these spectra have systematically different stellar parameters than HERMES spectra when measured with EPIC.
We identified 18 target stars that have been observed with both HERMES in GALAH DR2 as well as with HARPS in the ESO phase 3 database\footnote{\href{http://archive.eso.org/wdb/wdb/adp/phase3\_spectral/form}{http://archive.eso.org/wdb/wdb/adp/phase3\_spectral/form}}.
We analysed both the HERMES and the HARPS spectra for these target stars and compared their measured stellar parameters with each other. The resulting stellar parameters and their differences are visualised in \Fref{fig:harps_hermes}.

We found that the weighted average of GALAH-HARPS differences are $\Delta (\Teff, \log g, $[Fe/H]$) = (-10.9 \pm \SI{10.6}{\kelvin}, -0.021 \pm \SI{0.018}{dex}, -0.025 \pm \SI{0.007}{dex})$. These systematic effects can explain the differences that we see between stellar parameters calculated for the solar atlas and the HARPS solar spectra observed from reflected light (\Sref{sec:solar_test}). 
The most likely source for these differences is that HARPS spectra are not corrected for telluric features while both GALAH spectra and the solar atlas are corrected for telluric absorption/emission. We expect this to influence the normalisation process of EPIC (\Sref{sec:Precise_renorm}). We use HARPS spectra to re-calibrate EPIC with a `higher-order correction' in \Sref{sec:High_precision} so there is the possibility of small systematic effects in the stellar parameter measurements. However, only the linear and higher-order terms of this correction are used, so this should be a very small effect, while the `reference-point correction' is based on the solar atlas which means that these systematic effects would be reduced, especially for solar analogue spectra.

\subsection{Sun-like stars: higher-order correction}\label{sec:High_precision}
\begin{figure*}
	\centering
	\includegraphics[width=0.9\textwidth]{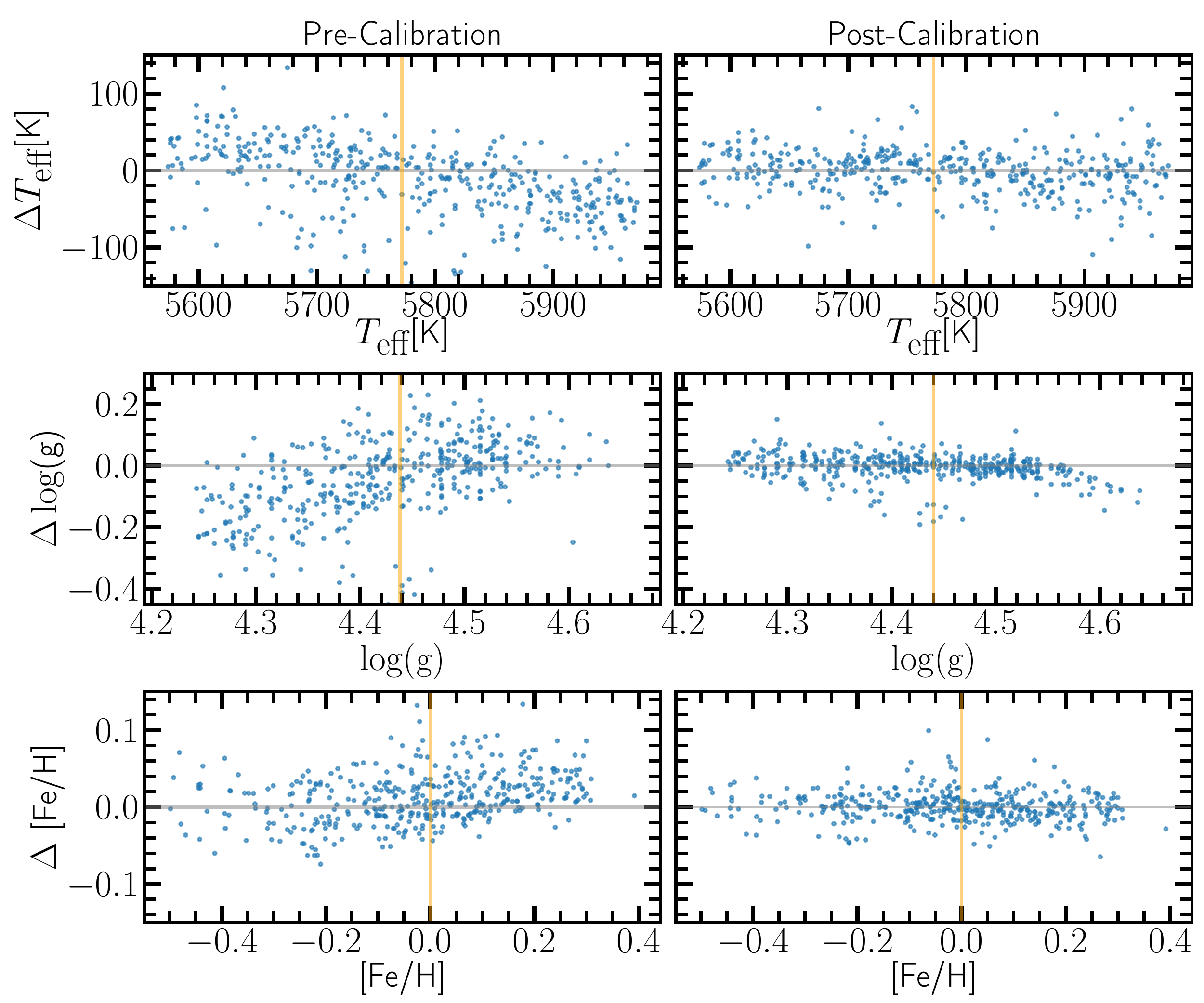}
	\caption{EPIC stellar parameter measurement compared to the analysis of \citet{Casali2020}. The panels from top to bottom show the effective temperature, surface gravity (g in cm/s$^2$) and metallicity. The $y$-axis represents the difference between stellar parameters calculated with EPIC and those of \citet{Casali2020}. The $x$-axis shows the parameter as measured by \citet{Casali2020}. The orange vertical lines represent solar values. The left panels show the differences before the `higher-order correction' and the right panels shows the improvement after we apply it.}
	\label{fig:casali_test}
\end{figure*}

After calibrating solar stellar parameters in EPIC we test the algorithm on a wider range of stars. Since we aim to minimise systematic errors for solar analogue stars (\Eref{eq:SA_def}) we need to find such targets in the literature to perform this test.
\citet{Casali2020} provided high precision stellar parameter estimations for 560 Sun-like targets observed with the HARPS spectrograph. The uncertainties calculated in this analysis are small ($\sigma(\Teff)=\SI{10}{\kelvin}$, $\sigma(\log g)=\SI{0.03}{dex}$, $\sigma(\textrm{[Fe/H]})=\SI{0.01}{dex}$) and the calibration of $\Teff$, $\log g$ and [Fe/H] was done using the modelling process of the q2 algorithm \citep[][]{Ramirez2014}.
This provides an independently observed and analysed set of stars with well-determined stellar parameters which allows us to test EPIC's stellar parameter measurements. We gathered 458 spectra from HARPS for targets previously used in \citet{Casali2020} from the ESO phase 3 database. This set of spectra was selected only by their availability in the archive, so we do not expect additional systematic effects from the fact that not all 560 stars were used.
We applied the EPIC algorithm, including the `reference-point correction', to these available data.

In the left three panels of \Fref{fig:casali_test} we can see how stellar parameters from the EPIC analysis compare to the higher-precision, independent values from \citet{Casali2020}. These panels show systematic differences in $\Teff$ and $\log g$ that vary approximately linearly with the value of these parameters, while [Fe/H] seems to vary slightly non-linearly. These systematic effects are visible because our initial calibration, based on The Cannon measurements, differs from the calibration of \citet[][]{Casali2020}.
In order to match the stellar parameter calibration of \citet[][]{Casali2020}, we need to correct for systematic effects that vary with respect to stellar parameters: we call this process a `higher-order correction'.
We re-calibrate by fitting a polynomial of the following form to each of the stellar parameters, $X$:
\begin{gather}
    dX = a_{1,X} \delta \Teff + a_{2,X} \delta\log g + a_{3,X} \textrm{[Fe/H]} + a_{4,X} \textrm{[Fe/H]}^2 , \label{eq:higher_order2}
\end{gather}
where $dX$ are the stellar parameter deviations between EPIC and \citet{Casali2020}, $\delta \Teff$ and $\delta\log g$ are the effective temperature and surface gravity differences with respect to the solar values (i.e.\ $\delta \Teff = \Teff - \SI{5772}{\kelvin}$ and $\delta \log g = \log g - 4.438$) and $a_{i,X}$ are the fit parameters in the calibration process. This polynomial is subtracted from the measured stellar parameters to arrive at the final calibration of EPIC.

The right panels of \Fref{fig:casali_test} visualise the effect of this process when compared to the left panels. They show that EPIC's stellar parameters (post-calibration) match the \citet[][]{Casali2020} values within a small uncertainty range. Additionally, the scatter seen in the left panels is substantially reduced in the right panels. This is because the systematic effects described are not just dependent on the stellar parameter they are affecting, e.g.\ systematic effects in $\Teff$ are also influenced by [Fe/H]. The scatter is reduced because these dependencies are addressed in \Eref{eq:higher_order2}.
Therefore, we conclude that the systematic differences between our measured stellar parameters with EPIC and those from \citet{Casali2020} have been calibrated out, with a possible exception at higher $\log g$ values (i.e.\ $\log g>\SI{4.54}{dex}$) which we need to be cautious of.
Since EPIC's calibration is fundamentally tied to the analysis of \citet[][]{Casali2020} it might need to be revisited in the future as other works improve their stellar parameter calibrations.

\subsection{EPIC and GALAH/The Cannon}\label{sec:Cannon}
\begin{figure*}
	\centering
	\includegraphics[width=0.9\textwidth]{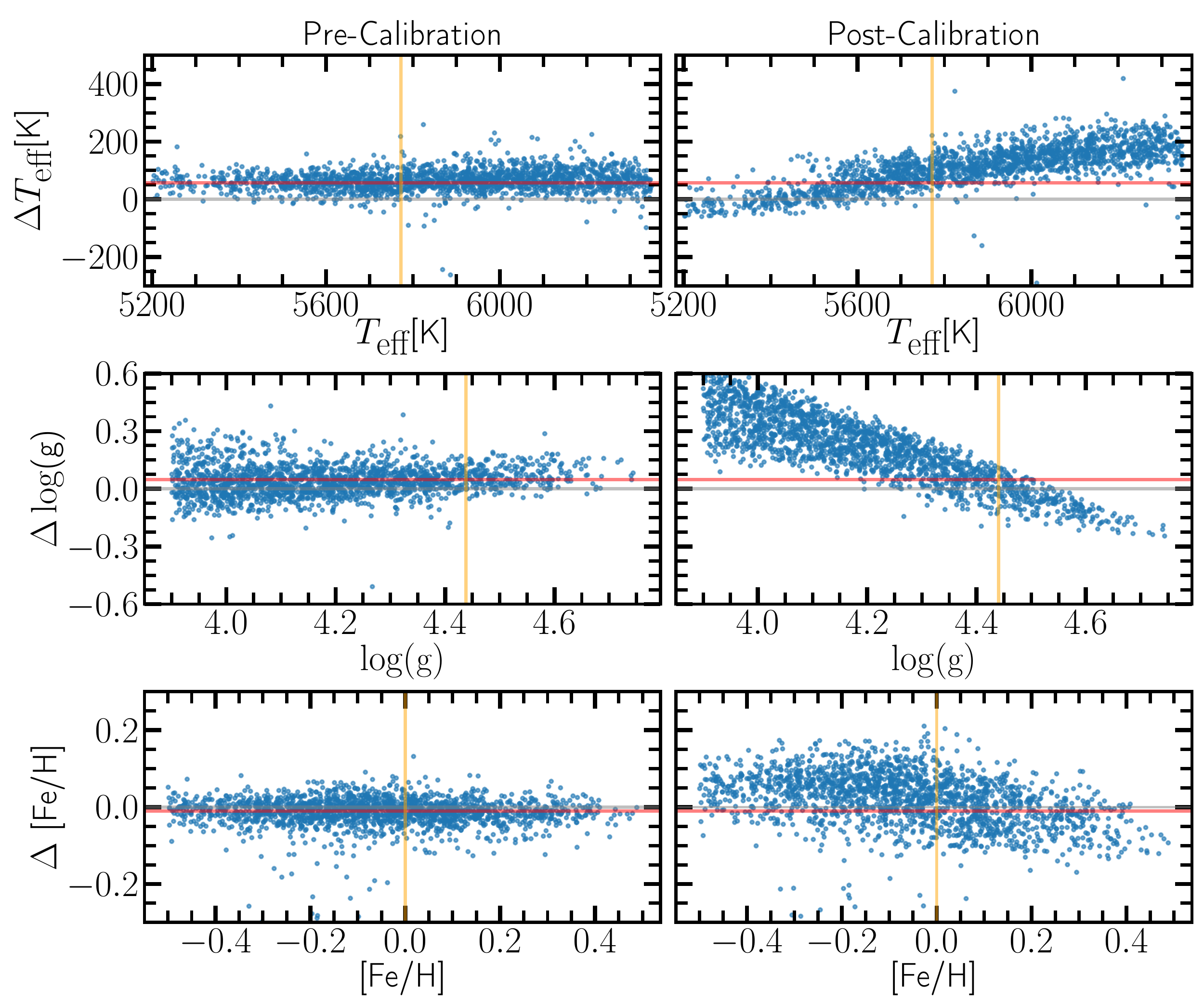}
	\caption{EPIC stellar parameter measurement compared to the values provided in GALAH DR2 calculated with The Cannon \protect\citep[][]{cannon2015}. The panels from top to bottom show the effective temperature, surface gravity (g in cm/s$^2$) and metallicity. The $y$-axes show the difference between EPIC's stellar parameters and The Cannon's. The $x$-axes show the parameters as measured with The Cannon. The vertical orange lines represent the solar values. The red horizontal line highlights the `reference-point correction' that has been applied to EPIC's stellar parameters in all panels. The left panels show the differences before the `higher-order calibration' while the right panels show them after that is applied.}
	\label{fig:GALAH_test}
\end{figure*}

EPIC currently focuses on measuring stellar parameters from HERMES spectra, so it is natural to compare its results with the GALAH stellar parameter measurements derived from HERMES spectra using The Cannon \citep[][]{cannon2015}. EPIC was initially calibrated to The Cannon stellar parameters because the stacked spectra \citep[][]{Zwitter2018} combine many GALAH DR2 spectra whose parameters were calculated with The Cannon. Therefore, we expect pre-calibration stellar parameters from EPIC to agree with stellar parameters in GALAH DR2.

We analysed a sub-set of GALAH DR2 with EPIC. These spectra were selected from GALAH to have measured stellar parameters in the following ranges: $\Teff=[5200, 6400] \SI{}{\kelvin}$, $\log g=[3.8, 5.0]\SI{}{dex}$ and [Fe/H] $=[-0.5, 0.5]\SI{}{dex}$. Within these stellar parameter ranges, we randomly selected $\sim1{,}858$ spectra for a workable test data-set. In this analysis we focused on GALAH's second data release (DR2) \citep[][]{Buder2018} which is the basis for the stacked spectra \citep[][]{Zwitter2018} we use for the model creation (\Sref{sec:stellar_par}). \Fref{fig:GALAH_test} shows EPIC's stellar parameter values compared to those of the GALAH survey. We visualise EPIC's stellar parameter estimation before (left three panels) and after (right three panels) the `higher-order calibration'. The `reference-point correction' is done before this comparison and the red horizontal line visualises the shift it caused in our calibration. 
Therefore, we expect the left column of \Fref{fig:GALAH_test} to show only statistical fluctuations around the `reference-point correction' line while the values calibrated to \citet[][]{Casali2020} should show systematic effects between the two calibrations.

The effective temperature and metallicity show agreement between EPIC's and The Cannon's stellar parameters pre-calibration while the surface gravity seems to have a minor systematic effect which becomes most apparent at low and high $\log g$. A possible explanation for this systematic effect is that different choices for line lists affect the overall stellar parameter measurements, which is especially true for $\log g$. The dependence of absorption feature strengths -- and therefore the precision of our measurements -- on $\log g$ varies on a line-by-line basis, so different choices of line list for EPIC can lead to systematic effects as seen in \Fref{fig:GALAH_test}. However, these offsets are small and will be calibrated out when doing the `higher-order correction' so it is not a major concern. Line-by-line differences in stellar parameter dependence are present for $\Teff$ and [Fe/H], but different features generally follow the same qualitative trends for these parameters so these differences are negligible compared to those in $\log g$.
We conclude that our replication of GALAH DR2 stellar parameters works as intended for the uncalibrated version of EPIC. After the application of our `reference-point' and `higher-order correction' (right-hand side of \Fref{fig:GALAH_test}), we see systematic effects in all stellar parameters. There is a clear gradient for $\Teff$ and $\log g$ in their respective panels. Although that may seem concerning at first, a direct comparison between GALAH DR2 and stellar parameters from \citet[][]{Casali2020} yields similar results. The same logic applies to [Fe/H] although the systematic differences between GALAH DR2 and \citet[][]{Casali2020} appear more complex than a linear dependence.

\subsection{Uncertainties} 
\begin{figure}
    \centering
    \includegraphics[width=0.475\textwidth]{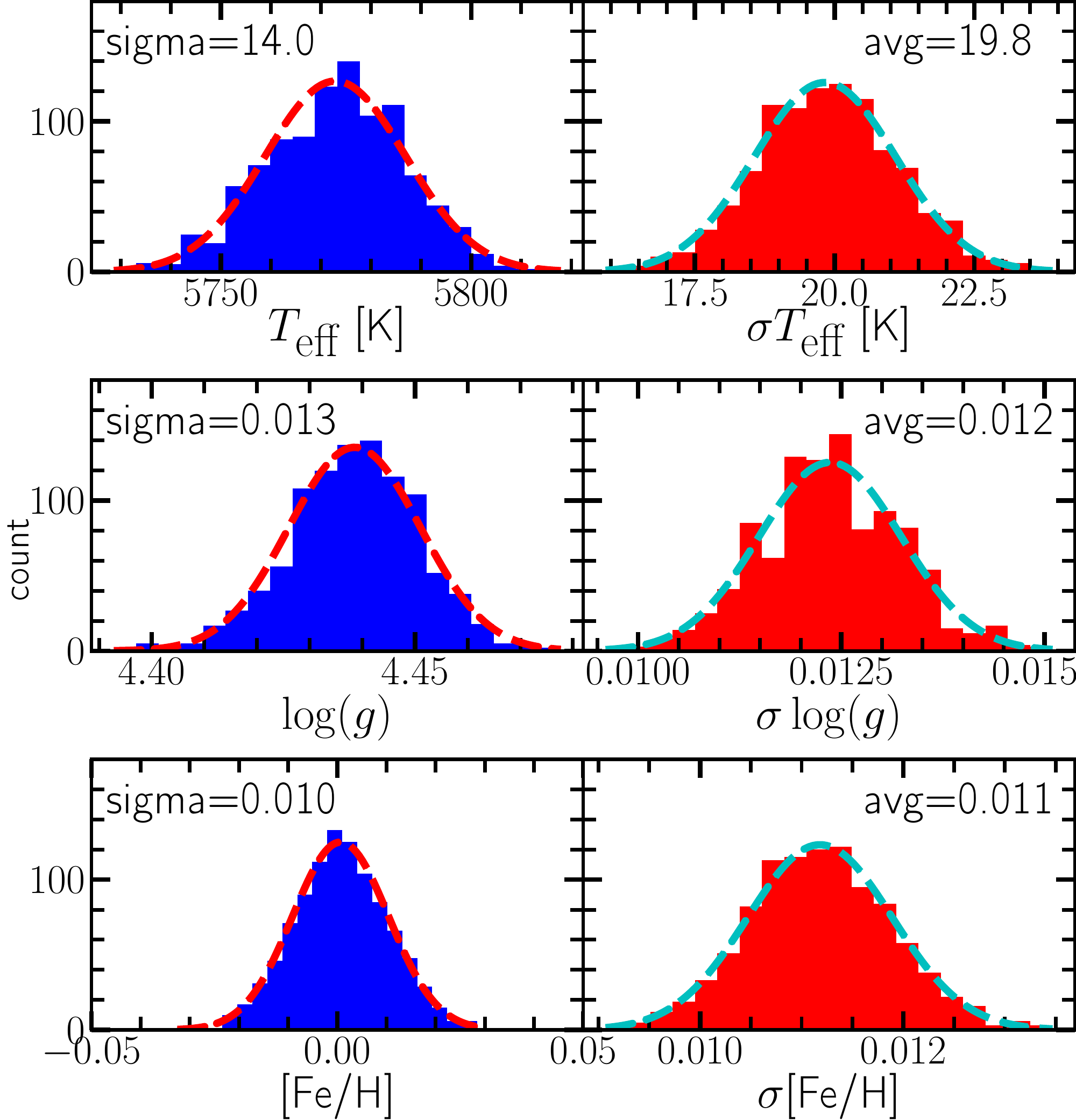}
    \caption{EPIC analysis for the solar atlas with random noise corresponding to $S/N\approx50$ in the red CCD of HERMES. The histograms show EPIC's stellar parameter measurements for the solar atlas \citep[][]{Chance2010} with simulated uncertainties for $1{,}000$ realisations. The left panels show the distribution of the resulting stellar parameters with a Gaussian overlay that uses standard deviation and average of the data. The right panels show the corresponding stellar parameter uncertainties that EPIC estimates for these spectra. Note the good correspondence between the sigma/dispersion in the left panel and the average calculated uncertainty in the right panels.}
    \label{fig:MC_test}
\end{figure}

\begin{figure}
    \centering
    \includegraphics[width=0.445\textwidth]{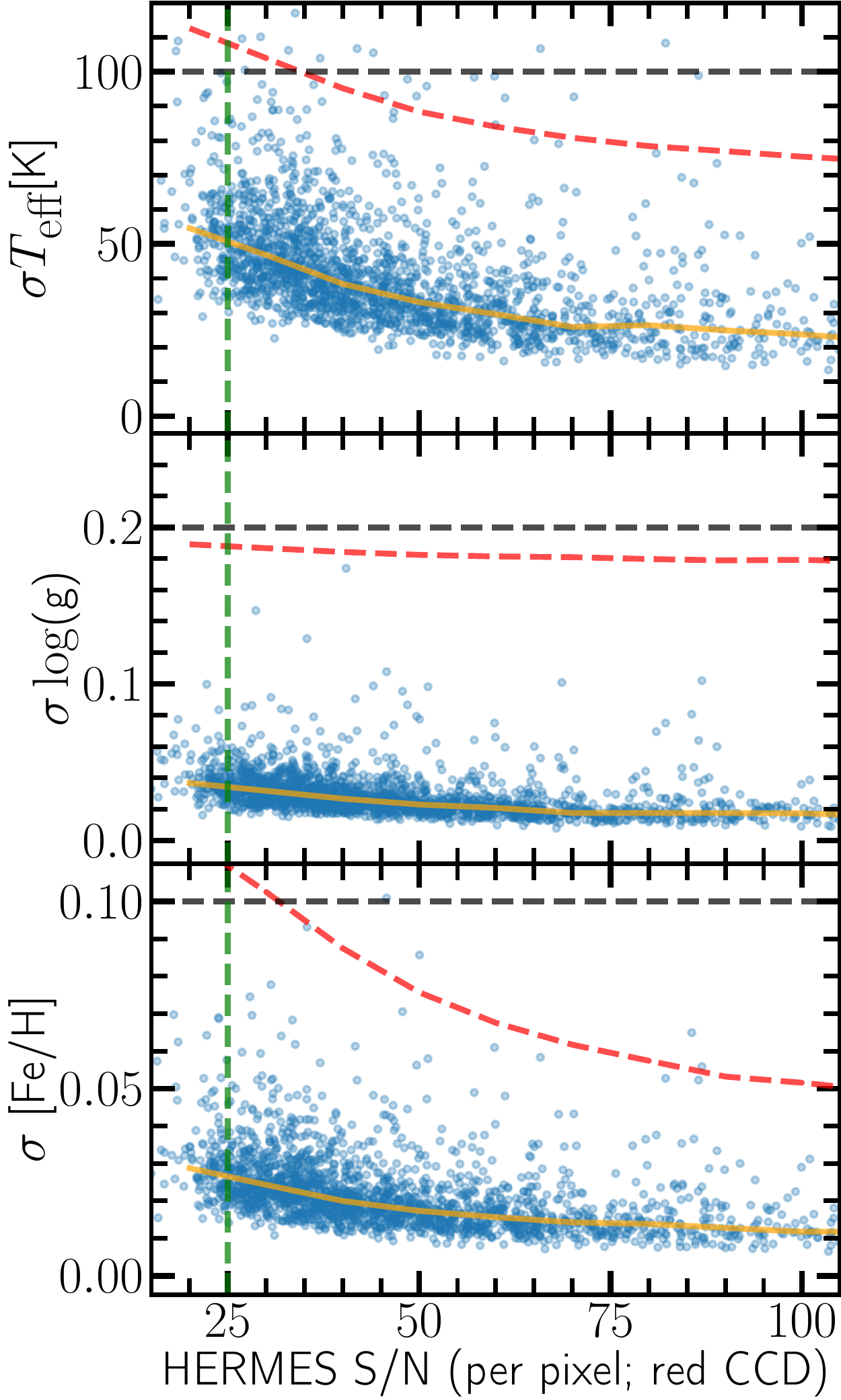}
    \caption{Statistical uncertainties for stellar parameters measured with the EPIC algorithm. The black dashed horizontal line represents our maximum uncertainty requirement at $S/N=25$ (green dashed vertical line) which is derived from our solar analogue definition. The blue data points represent the uncertainties EPIC calculated for HERMES spectra and the orange line represents the average uncertainty as a function of $S/N$. The red dashed line represents the average uncertainty for GALAH/The Cannon stellar parameters.}
    \label{fig:uncertainties}
\end{figure}

An important measure of how well EPIC measures stellar parameters is the total statistical uncertainties it derives for $\Teff$, $\log g$ and [Fe/H].
The sources of these uncertainties are the statistical errors in the flux and continuum level errors which lead to uncertainties in the EW, and modelling uncertainties such as deviations of the measured EWs from those our model would predict, and uncertainties in the model parameters ($c_i$ in \Sref{sec:stellar_par}).
The final uncertainties are computed using the resulting variances in the stellar parameter fit process (\Sref{sec:stellar_par}).

We tested if the measured uncertainties from EPIC are reasonable by applying random noise (corresponding to $S/N\approx50$) to the solar atlas \citep[][]{Chance2010}, which has been prepared as a target spectrum (\Sref{sec:high_res_prep}), and analysing it with the algorithm. We performed $1{,}000$ realisations of this simulated solar spectrum to check that the variance in the results matched the uncertainties predicted by EPIC. The left panels of \Fref{fig:MC_test} shows the stellar parameters as measured by EPIC with a Gaussian overlay that is created from the data average and standard deviation. The right panels show the distribution of calculated uncertainties for the same data with another Gaussian overlay. EPIC's measured stellar parameters follow a distribution that can be approximated by a Gaussian curve as shown by the overlay function in the left panels. The mean of the distribution is at solar stellar parameters which leads us to conclude that a moderate amount of noise does not cause significant systematic effects for measured stellar parameters.
\Fref{fig:MC_test} also shows that the standard deviation in the stellar parameters over the $1{,}000$ realisations is very similar to EPIC's uncertainty estimates on average. One possible exception is that EPIC may slightly overestimate the $\Teff$ uncertainties. Nevertheless, we can conclude that EPIC does not underestimate the stellar parameter uncertainties, so they are robust in this sense.

Additionally, we applied a reduced $\chi^2$ test to the stellar parameter differences between EPIC's analysis and that of \citet[][]{Casali2020}. We concluded that systematic effects have mostly been corrected (right panels of \Fref{fig:casali_test}) which means the reduced $\chi^2$ of the results should test how well our uncertainties match the variations in the residuals.
We found that EPIC possibly underestimates the uncertainties for $\Teff$, $\log g$ and [Fe/H] slightly ($\chi^2 = 1.25$, $\chi^2 = 1.32$ and $\chi^2 = 1.09$ respectively).

We determine the quality of EPIC's analysis by measuring the uncertainties for HERMES spectra with respect to their $S/N$. The data-set used in \Sref{sec:Cannon} is a sub-set of GALAH DR2 which spans $S/N\approx[20, 100]$ in the red CCD. We visualise the measured uncertainties for this data-set in \Fref{fig:uncertainties}.
The uncertainties need to be sufficiently low to allow the identification of solar twins/analogues with reasonable confidence. This motivated our maximum uncertainty limits (black dashed horizontal lines): $\sigma\left(\Teff, \log g, \textrm{[Fe/H]}\right) < \left(\SI{100}{\kelvin, \SI{0.2}{dex}, \SI{0.1}{dex}}\right)$. To search for distant solar twins/analogues, we need to have uncertainties lower than our set limits at $S/N\approx 25$ (green dashed vertical line), which is possible to achieve with the HERMES spectrograph for solar like stars at a distance of $\approx\SI{4}{kpc}$. As seen in \Fref{fig:uncertainties} the average uncertainties we measure with EPIC are below the set goals for all stellar parameters even at low $S/N$. We also note that our average uncertainties improve on those from The Cannon by at least a factor of 2 across the plotted range of S/N.

\subsection{Processing time}\label{sec:Process_time}
We have tested the processing speed of the algorithm with HERMES spectra from GALAH DR2. EPIC operates quickly and can measure stellar parameters for a single HERMES spectrum in an average time of $\approx10$\,sec on a 2.6\,GHz  Intel i7 Processor.
This means that it can easily be applied to very large data-sets, e.g.\ GALAH DR3 \citep[][]{Buder2021} -- which contains $\sim600{,}000$ HERMES spectra -- could be analysed in only $\sim1{,}600$ core-hours.

\section{Discussion}\label{sec:discussion}
In this section we discuss the potential impact of the EPIC method and compare it with other algorithms that measure stellar parameters. We also give a brief outlook on potential future applications of EPIC.

\subsection{Mapping \texorpdfstring{$\boldsymbol{\alpha}$}{alpha} throughout the Milky Way}
EPIC enables the identification/confirmation of solar analogue and twin stars with spectral data from HERMES. As mentioned earlier (\Sref{sec:Intro}) we can use these stars to probe the fine-structure constant $\alpha$. Berke et al. (2022a; in prep) and Murphy et al. (in prep) demonstrate that the separation between selected pairs of lines can be measured with $~\SI{5}{\metre\per\second}$ accuracy using solar analogues, allowing the fine-structure constant to be measured with a precision of $\Delta\alpha/\alpha \sim 10^{-7}$ from HARPS spectra of two or more stars. The ESPRESSO (Echelle SPectrograph for Rocky Exoplanets and Stable Spectroscopic Observations) instrument at the VLT (Very Large Telescope, $\SI{8.2}{\metre}$) is the successor of HARPS and should provide measurements in distant solar analogues with systematic errors as low as HARPS has.
ESPRESSO is highly stabilised, provides repeatable and precise wavelength calibrations and is mounted on the VLT which makes it possible to observe our faintest solar analogue targets. The maximum distance to which we can observe solar analogues is determined by the available ESPRESSO time and the required spectral quality ($S/N\gtrsim40$ per $\SI{0.4}{\kilo\metre\per\second}$ pixel) for each twin. Assuming eight hours of exposure time per star and negligible extinction we arrive at a brightness limit of $G\sim\SI{17.4}{mag}$ for ESPRESSO, which corresponds to a solar analogue at a distance of $\sim\SI{4}{kpc}$.

The target stars for an ESPRESSO observation need to be discovered beforehand, which we plan to do within our program to observe $\approx550$ solar analogue candidates with AAT/HERMES (Lehmann et al. in prep, Liu et al. in prep). From those targets we will choose the most Sun-like stars for ESPRESSO follow-up.
With AAT/HERMES, the faintest targets ($G\sim\SI{17.4}{mag}$) can be observed in $\sim18$ hours of exposure to reach $S/N\sim25$ per pixel.

In this paper we demonstrated that the EPIC algorithm can be applied to these low $S/N$ spectra with sufficient precision to identify solar analogue stars. Therefore, the algorithm is a key part of the preparation for ESPRESSO observations by providing target stars to the campaign.

\subsection{Comparison with other methods}\label{sec:Comparison_other_methods}
In this section we discuss the advantages and limitations of the EPIC algorithm with respect to other prominent spectroscopic methods for stellar parameter determination.
For this purpose, we selected the Spectroscopy Made Easy (SME) package \citep[][]{Piskunov2016}, The Cannon \citep[][]{cannon2015} as used for HERMES spectra in GALAH DR2 \citep[][]{Buder2018}, as well as the qoyllur-quipu (hereafter `q2') \citep[][]{Ramirez2014} algorithm used in \citet[][]{Casali2020}. We present a short summary of each method's functionality, and then discuss their intended applications, accuracy/systematic errors and statistical uncertainties in comparison to EPIC.

\subsubsection{SME tool}
The SME tool operates on specified spectral regions and compares them with synthetic spectra to derive the best fit stellar parameters. Using atomic/molecular line data, SME creates model spectra from first guesses for stellar parameters. It then varies the stellar parameters to minimise differences between the synthetic and observed spectrum.
This is in direct contrast to the methodology of EPIC: SME models the stellar astrophysics to produce synthetic spectra while EPIC derives stellar parameters in a fully phenomenological way.
Note that the stellar parameters of the spectral training set for EPIC were produced from The Cannon which leverages the astrophysical modelling of SME. So EPIC's initial stellar parameter measurements (before re-calibration; \Sref{sec:High_precision}) are tied to SME's approach.

\subsubsection{The Cannon}
The Cannon utilises a training set of spectra, with known stellar parameters, to model the behaviour of the (continuum normalised) flux in every pixel of the target spectrum with respect to stellar parameters.
The model created contains probability density functions for each continuum normalised pixel in a potential target spectrum. It varies the stellar parameters for the model to minimise the differences between model-predicted and measured flux to find solutions for $\Teff$, [Fe/H] and $\log g$.
In the case of GALAH DR2, The Cannon was trained on 10605 HERMES spectra for which stellar parameters have been calculated with the SME package.

The purpose of The Cannon, adopted by GALAH DR2, is to efficiently find stellar parameters for their spectral data-set of $342{,}682$ target stars. This can be achieved because, after the initial model training, the algorithm can derive stellar parameters for HERMES target spectra very efficiently and quickly.
The differences between EPIC and The Cannon stem from different final calibrations (a SME training set for The Cannon and a q2/MOOG training set for EPIC) and EPIC's measurement of EWs for individual absorption features, in contrast to the full spectrum modeling approach for The Cannon. Their shared purpose of efficient analysis for a large number of spectra makes The Cannon and EPIC the most closely comparable of the methods presented here.

\subsubsection{q2 algorithm}\label{sec:q2}
The q2 algorithm \citep[][]{Ramirez2014} was employed by \citet[][]{Casali2020} whose HARPS spectra and stellar parameter measurements we use for the final calibration of EPIC (\Sref{sec:High_precision}).
Similar to EPIC, q2 requires the EWs of selected Fe features, measured in a differential way, relative to a solar spectrum as input.
q2 makes use of the Kurucz (ATLAS9) grid of model atmospheres \citep[][]{Castelli2003} as well as MOOG 2014 \citep[][]{Sneden1973}.
It starts at literature values for stellar parameters and then iteratively changes them to minimise correlations in the residuals between stellar parameters and relative EW as well as stellar parameters and excitation potential. Additionally, they minimise the difference between Fe abundance measurements from FeI and FeII lines.
\citet[][]{Casali2020} employed the q2 algorithm to find precise stellar parameters and abundances for 560 solar twin candidate stars that have been observed with the HARPS spectrograph.
The q2 algorithm is comparable to EPIC as it uses a similar differential approach to the EW measurement and stellar parameter measurement. The main difference is that q2 determines stellar parameters using atomic line data and atmospheric models while EPIC's calibration of stellar parameters derives from its input stellar library.

\subsubsection{Comparing other algorithms with EPIC}
In comparison to the three methods above, our main aim for EPIC is the measurement of precise stellar parameters for solar analogue stars using spectroscopic data from HERMES.
The differences in intended use, as well as quality of data, are important to keep in mind when comparing the methods' systematic errors and statistical uncertainties, e.g.\ we would expect q2 in \citet[][]{Casali2020} to have low statistical uncertainties because they make use of high $S/N$, high resolving power data.
Furthermore, we note that both q2 and SME use models of stellar atmospheres to measure stellar parameters while The Cannon and EPIC are fundamentally phenomenological, basing their measurements solely on the stellar parameters embedded in their training sets.

\subsubsection{Systematic errors}
Systematic errors are a concern for each of these methods and minimising them is an important goal. In EPIC's case we need to minimise systematic errors to ensure that we can identify solar analogues at large distances (i.e.\ using low $S/N$ spectra).
Comparing physical quantities with other algorithms provides insight into how accurately we measure stellar parameters with EPIC. In \Sref{sec:High_precision} and \ref{sec:Cannon} we compared our measured stellar parameters directly to those derived by The Cannon (GALAH DR2) and q2 \citep[][]{Casali2020}. By applying different calibration corrections, EPIC is able to match both q2's and The Cannon's stellar parameters within statistical uncertainties. However, we saw that any one calibration of EPIC did not match the results of both algorithms, which leads us to the conclusion that systematic differences between q2 (based on MOOG) and The Cannon (trained on SME results) exist. These differences likely result from the different physics employed in the stellar atmospheric models or the differences in their algorithms and/or training sets.

We decided to base our final calibration in this work on the results of \citet[][]{Casali2020} for two reasons: (i) this calibration provides measurements for solar spectra that are consistent with solar stellar parameters (see \Fref{fig:casali_test}); and (ii) \citet[][]{Casali2020} tested their surface gravity results against photometric surface gravity results and found them to be consistent.

\subsubsection{Statistical uncertainties}
The precision of each method, i.e.\ the statistical uncertainties in stellar parameters, is another important measure of their effectiveness. Our main motivation to create EPIC was to reduce the statistical uncertainties for low $S/N$ spectra to the point where we can identify solar twins and analogues with high confidence. The statistical uncertainties for each method are driven by two different factors. The first is statistical noise, i.e.\ the photon statistical flux uncertainties in the spectral data and errors in the continuum level.
The second contribution comes from model uncertainties which includes both deviations of individual spectra from the model predictions and uncertainties in the model parameters themselves. In EPIC's case these are the $c_i$ model parameters in \Sref{sec:stellar_par}. 

At our target $S/N$ for HERMES spectra, these two sources of statistical uncertainties produce values far below our set budget for EPIC ($\approx25$ in the red CCD, corresponding to the vertical green dashed line in \Fref{fig:uncertainties}). At high $S/N$ we can see that there is an uncertainty floor which is caused by the modelling errors: $\sigma(\Teff, \log g, \textrm{[Fe/H]})=(\SI{20}{\kelvin}, \SI{0.04}{dex}, \SI{0.014}{dex})$. 
Note that the uncertainties for the solar atlas in \Sref{sec:solar_test} are smaller than this uncertainty floor because in that case we are using the reference spectrum as if it was a target spectrum. This reduces some sources of scatter in the EW measurements. For example, the treatment of the continuum in target and reference is identical, and there will be only very small errors in the radial velocity corrections.

We can compare EPIC's uncertainties to those of The Cannon (GALAH DR2, see \Fref{fig:uncertainties}).
On average, EPIC produces $\lesssim 2-3$ times smaller uncertainties than The Cannon for the same spectra in GALAH DR2.
We attribute this to EPIC's differential approach, in contrast with The Cannon which attempts to measure absolute values for the stellar parameters. EPIC's advantage of having a solar spectrum as a comparison provides a consistent means of estimating radial velocities and continua. This reduces the statistical and systematic error in the estimate of an individual feature's EW. As a result, the stellar parameters have reduced statistical uncertainties.

The quoted uncertainties for the 560 target solar twin candidates in \citet[][]{Casali2020} that were derived with the q2 algorithm are: $\sigma(\Teff, \log g, \textrm{[Fe/H]})=(\SI{10}{\kelvin}, \SI{0.03}{dex}, \SI{0.01}{dex})$. The flux noise of the spectra are negligible, so these numbers almost entirely represent their modelling uncertainties.
By means of comparison, for a high $S/N$ HERMES spectrum ($\gtrsim100$) the average uncertainties from EPIC are: $\sigma(\Teff, \log g, \textrm{[Fe/H]})=(\SI{22}{\kelvin}, \SI{0.04}{dex}, \SI{0.014}{dex})$.
Although EPIC's uncertainties are larger than those quoted in \citet[][]{Casali2020} for high $S/N$ spectra, the fact that EPIC's are derived from low resolving power HERMES spectra shows how effective the differential approach is.

\subsubsection{Final remarks}
EPIC's advantages in comparison with these other methods are that its differential approach minimises statistical and systematic uncertainties for stellar parameters, it can produce robust results even for low $S/N$ spectra, and operates efficiently with short computation times (\Sref{sec:Process_time}). EPIC therefore proves to be the best means for identifying distant solar twins/analogues with low resolution HERMES spectra. However, the EPIC algorithm is fundamentally limited as it uses data from other methods as a training set and will only provide reliable results for target stars within the same stellar parameter range defined by the training set.

\section{Conclusions}\label{sec:conclusion}
We presented the novel EPIC algorithm to measure spectroscopic stellar parameters ($\Teff$, $\log g$ and [Fe/H]) of target stars with a focus on solar twin and analogue candidates. A high precision reference spectrum -- the solar atlas from \citet[][]{Chance2010} -- is needed to optimise the precision of this fully differential method. We measured the EWs of absorption features within HERMES spectra from GALAH with respect to the reference and spent considerable effort to minimise the systematic effects: precise wavelength alignment between absorption features of reference and target spectra, matching of resolving powers, and re-normalisation to a shared continuum.
Using the spectral library by \citet[][]{Zwitter2018} we created a model that connects differences in EWs between reference and target spectra to stellar parameter differences.

We analysed GALAH DR2 spectra of target stars with known stellar parameters to test the accuracy and capabilities of the method and found an offset between EPIC's measured stellar parameters and literature values for the same stars, e.g.\ solar stellar parameters had a measured offset of $\Delta (\Teff, \log g$, [Fe/H]$) = (\SI{61}{\kelvin}, \SI{0.06}{dex}, -\SI{0.004}{dex})$ (\Sref{sec:solar_test}). The most likely cause for these offsets is initial miscalibrations in The Cannon \citep[][]{cannon2015} which EPIC inherits from the training set (\Sref{sec:stellar_par}). \citet{Buder2018} found the same mean temperature offset between GALAH (DR2) and indepdendent photometric estimates.
We re-calibrated EPIC to match stellar parameters of 1) the Sun, using a `reference-point-correction' which applies constant stellar parameter corrections, and 2) solar analogue stars,  using a `higher-order-correction' which applies a stellar parameter dependent correction.
The measurements of EPIC are based on a training set \citep[][]{Zwitter2018} that uses the stellar parameter measurement algorithm The Cannon. After the corrections, EPIC's final calibration matches the stellar parameters measured in \citet[][]{Casali2020}, which is based on the q2 algorithm, it more reliably predicts solar stellar parameters and agrees with photometric surface gravity measurements. In these ways, EPIC is dependent on the calibration of other stellar parameter measurement algorithms which do not agree with each other, so it might be necessary to revisit EPIC's calibration in the future.

We have tested the measured uncertainties of the EPIC algorithm and found good correspondence with the scatter in measured quantities. Our goal with EPIC is to minimise these uncertainties in order to gain confidence when identifying solar analogue stars. These uncertainties are significantly lower than those found when measuring stellar parameters with The Cannon \citep[][]{cannon2015}, e.g. EPIC has an average temperature uncertainty of $\sigma \Teff\sim\SI{50}{K}$ while the Cannon has $\sigma \Teff\sim\SI{110}{K}$ at $S/N=25$. These low average uncertainties, i.e.\ $\sigma(\Teff, \log g, \textrm{[Fe/H]}) = (\SI{50}{\kelvin}, \SI{0.8}{dex}, \SI{0.3}{dex})$ at $S/N=25$, allow us to identify solar twins (\Eref{eq:SA_def}) with up to $92.9\%$ confidence and solar analogues (\Eref{eq:SA_def}) with up to $98.5\%$ confidence. The uncertainties of EPIC also compare well with those derived from high $S/N$, high resolving power spectra from \citet[][]{Casali2020}. For HERMES spectra with high $S/N$, EPIC's uncertainties are only higher by a factor of 1.4 in [Fe/H] and 2.2 in $\Teff$, which is encouraging given the comparatively low resolving power and wavelength coverage of HERMES compared to HARPS.

The combination of short processing time ($\lesssim \SI{10}{\second}$ per spectrum), reasonable precision of stellar parameters with limited spectral information from HERMES, and systematic effects that can be minimised by re-calibration, are what makes the EPIC algorithm valuable for stellar identifications of Sun-like stars.

We can extend our identification of solar analogues to the most distant ($\sim\SI{4}{kpc}$) solar candidate stars by observing photometrically pre-selected targets (Liu et al. in prep) with HERMES and measure their stellar parameters with EPIC.
These solar analogue stars can be used to measure the fine structure constant $\alpha$ and map possible variations of it across the galaxy (Murphy et al. in prep, Berke et al. 2022a, b; in prep).
Additionally, we intend to apply the EPIC algorithm to the third data release of the GALAH survey \citep[][]{Buder2021} to identify a large number of solar analogues. This will potentially identify thousands of new solar analogues out to distances of $\sim\SI{1}{kpc}$ compared to the few hundred within $\sim\SI{800}{pc}$ known today.
These analogues will be an important, relatively homogeneous sample for studies of chemical evolution, stellar evolution and galactic archaeology.

\section*{Acknowledgements}
We thank Janez Kos for providing resolving power maps of the HERMES spectrograph, and Karin Lind for providing the line list which was used in GALAH DR2. 
We also thank the anonymous referee for suggestions which improved the clarity of the paper.
CL, MTM and FL acknowledge the support of the Australian Research Council through \textsl{Future Fellowship} grant FT180100194.
We used Python 3.7.4 in the EPIC software and would like to thank the Python Software Foundation \citep[][]{python}.
Furthermore, we want to acknowledge our usage of the Python software packages Astropy \citep[][]{astropy}, Matplotlib \citep[][]{matplotlib}, Numpy \citep[][]{numpy} and Scipy \citep[][]{scipy}.

\section*{Data availability}
Spectra from the GALAH survey are available at \href{https://datacentral.org.au/services/download}{https://datacentral.org.au/services/download} with instructions at \href{https://www.galah-survey.org}{https://www.galah-survey.org}. The GALAH DR2 median spectra can be accessed at \href{https://datacentral.org.au/teamdata/GALAH/public/GALAH\_DR2}{https://datacentral.org.au/teamdata/GALAH/public/GALAH\_DR2}. The solar atlas (KPNO2010) is published at \href{http://kurucz.harvard.edu/sun/irradiance2005}{http://kurucz.harvard.edu/sun/irradiance2005}.
HARPS data products can be obtained from \href{http://archive.eso.org/wdb/wdb/adp/phase3\_spectral/form}{http://archive.eso.org/wdb/wdb/adp/phase3\_spectral/form}. 
The EPIC code is made available at \href{https://doi.org/10.5281/zenodo.6029325}{https://doi.org/10.5281/zenodo.6029325}.



\bibliographystyle{mnras}
\bibliography{Bibliography} 

\newpage
\appendix

\section{Line List}
\begin{table*}
    \centering
	\begin{tabular}{cc|cc|cc} 
		\hline
       	Species & Wavelength [\AA] & Species & Wavelength [\AA] & Species & Wavelength [\AA] \\
       	\hline
       	\ion{Mg}{i} & 4731.35 & \ion{Fe}{i}* & 5703.13 & \ion{O}{i}* & 7777.53 \\
       	\ion{Fe}{i} & 4734.92 & \ion{Mg}{i} & 5712.67 & \ion{Ni}{i}* & 7791.07 \\
       	\ion{Mn}{i} & 4740.42 & \ion{Fe}{i} & 5733.35 & \ion{Ni}{i}* & 7799.73 \\
       	\ion{Fe}{i} & 4742.86 & \ion{Fe}{i}* & 5754.73 & \ion{Rb}{i}* & 7802.21 \\
       	\ion{Fe}{i} & 4747.13 & \ion{Ni}{i} & 5756.27 & \ion{Al}{i}* & 7837.46 \\
       	\ion{Ti}{i} & 4759.45 & \ion{Si}{i} & 5773.75 & \ion{Al}{i}* & 7838.29 \\
       	\ion{Ti}{i} & 4760.60 & \ion{Fe}{i} & 5776.68 & \ion{Si}{i}* & 7852.14 \\
       	\ion{Mn}{i} & 4762.84 & \ion{Cr}{i} & 5785.47 & UL & 4723.47 \\
       	\ion{Ti}{ii} & 4765.86 & \ion{Cr}{i} & 5789.52 & UL & 4745.74 \\
       	\ion{Fe}{i} & 4780.78 & \ion{Si}{i} & 5794.68 & UL & 4749.47 \\
       	\ion{Fe}{i} & 4789.17 & \ion{Fe}{i} & 5808.33 & UL & 4765.26 \\
       	\ion{Fe}{i} & 4790.10 & \ion{Fe}{i} & 5810.83 & UL & 4774.16 \\
       	\ion{Cr}{i} & 4790.67 & \ion{Fe}{i} & 5853.84 & UL & 4791.00 \\
       	\ion{Ti}{ii}* & 4799.87 & \ion{Ca}{i} & 5859.07 & UL & 4808.33 \\
       	\ion{Cr}{i} & 4802.37 & \ion{Fe}{i}* & 5861.21 & UL & 4814.84 \\
       	\ion{Fe}{i} & 4804.22 & \ion{Fe}{i} & 5863.98 & UL & 4819.16 \\
       	\ion{Fe}{i} & 4809.49 & \ion{Ca}{i}* & 6495.58 & UL & 4825.50 \\
       	\ion{Zn}{i} & 4811.88 & \ion{Fe}{i}* & 6496.78 & UL & 4830.74 \\
       	\ion{Ti}{i} & 4821.76 & \ion{Fe}{i}* & 6500.73 & UL & 4834.09 \\
       	\ion{Ni}{i} & 4832.53 & \ion{Fe}{ii}* & 6517.88 & UL & 4840.91 \\
       	\ion{Ti}{i} & 4842.23 & \ion{Fe}{i}* & 6520.17 & UL & 4849.61 \\
       	\ion{Fe}{i} & 4845.37 & \ion{Fe}{i}* & 6548.05 & UL & 4852.85 \\
       	\ion{Cr}{ii} & 4849.59 & \ion{Ni}{i}* & 6588.13 & UL & 5677.00 \\
       	\ion{Ti}{ii} & 4850.52 & \ion{Fe}{i} & 6594.73 & UL* & 5693.09 \\
       	\ion{Y}{ii} & 4856.22 & \ion{Fe}{i}* & 6595.69 & UL* & 5695.25 \\
       	\ion{Ti}{ii} & 4866.97 & \ion{Fe}{i} & 6599.38 & UL & 5707.59 \\
       	\ion{Ni}{i} & 4867.63 & \ion{Sc}{ii} & 6606.43 & UL & 5708.68 \\
       	\ion{Ti}{ii} & 4875.37 & \ion{Fe}{i} & 6610.94 & UL & 5716.69 \\
       	\ion{Fe}{i} & 4877.24 & \ion{Fe}{i} & 6629.37 & UL & 5719.45 \\
       	\ion{Fe}{i} & 4883.51 & \ion{Ni}{i} & 6645.48 & UL & 5753.66 \\
       	\ion{Y}{ii} & 4885.05 & \ion{Fe}{i} & 6679.83 & UL & 5792.62 \\
       	\ion{Cr}{ii} & 4885.97 & \ion{Al}{i} & 6697.87 & UL & 6501.47 \\
       	\ion{Fe}{i} & 4892.12 & \ion{Al}{i} & 6700.52 & UL* & 6571.05 \\
       	\ion{Fe}{i} & 4892.86 & \ion{Fe}{i} & 6705.42 & UL* & 6576.88 \\
       	\ion{Sc}{ii} & 5659.47 & \ion{Cr}{i} & 6723.70 & UL & 6706.96 \\
       	\ion{Fe}{i} & 5664.09 & \ion{Fe}{i} & 6727.21 & UL & 6728.57 \\
       	\ion{Si}{i} & 5667.13 & \ion{Fe}{i}* & 7712.49 & UL* & 7716.42 \\
       	\ion{Sc}{ii} & 5668.72 & \ion{Fe}{ii}* & 7713.84 & UL* & 7725.34 \\
       	\ion{Fe}{i} & 5669.09 & \ion{Ni}{i}* & 7717.71 & UL* & 7729.78 \\
       	\ion{Sc}{ii} & 5670.63 & \ion{Fe}{i}* & 7750.40 & UL* & 7744.87 \\
       	\ion{Fe}{i} & 5680.60 & \ion{Ni}{i}* & 7751.02 & UL* & 7782.75 \\
       	\ion{Na}{i}* & 5684.21 & \ion{Fe}{i}* & 7753.24 & UL* & 7810.04 \\
       	\ion{Na}{i}* & 5689.78 & \ion{O}{i}* & 7774.08 & UL* & 7834.36 \\
       	\ion{Si}{i}* & 5692.00 & \ion{O}{i}* & 7776.31 &  &  \\
        \hline
	\end{tabular}
	\caption{Line list used in EPIC. The wavelength is given in vacuum. The lines selected have previously been used by \protect\citet{Datson2015} and \protect\citet{Buder2018} while lines labelled `UL' (unknown line) have been added in this work as described in \Sref{sec:linelist}. Lines marked with a `*' are not used when analysing HARPS spectra.}\label{tab:linelist}
\end{table*}


\bsp	
\label{lastpage}
\end{CJK*}
\end{document}